\documentclass{article}

\usepackage{arxiv}
\makeatletter
\AtBeginDocument{%
  \pagestyle{fancy}%
  \rhead{}%
}

\renewcommand{\keywords}[1]{\par\noindent\textbf{Keywords:} #1}
\makeatother
\usepackage{booktabs}
\usepackage{tabularx}
\usepackage[utf8]{inputenc} % allow utf-8 input
\usepackage[T1]{fontenc}    % use 8-bit T1 fonts
\usepackage{amssymb} % for \varnothing
\usepackage{amsthm} % for proof environment

\usepackage{url}            % simple URL typesetting
\usepackage{amsfonts}       % blackboard math symbols
\usepackage{microtype}      % microtypography
\usepackage{graphicx}
\usepackage{amsmath}
\usepackage{array} 
\usepackage{natbib}
\usepackage{lineno}         % for \linenumbers
\usepackage{longtable}
\usepackage{pdflscape}

\usepackage{tikz}
\usepackage{tikz-3dplot}
\usetikzlibrary{arrows.meta,positioning,calc,fit,shapes.geometric}
\definecolor{diagramblue}{RGB}{13,101,139}

\usepackage{pgfplots}
\usepgfplotslibrary{groupplots}
\pgfplotsset{compat=1.18}

\usepackage{enumitem}
\setlist[itemize]{leftmargin=1.6em}
\setlist[enumerate]{leftmargin=1.8em}

\usepackage{authblk}

\makeatletter
\renewcommand\AB@affilnote[1]{\textsuperscript{#1}\,}
\makeatother

\usepackage[colorlinks=true,linkcolor=black,citecolor=blue,urlcolor=blue]{hyperref}

\title{A Methodology for Auditable Trustworthiness Levels in AI Lifecycle Governance}

\author[1,2,3]{Andrea Ferrario%
\thanks{\href{mailto:aferrario@ethz.ch}{\texttt{aferrario@ethz.ch}}}%
\hspace{0.35em}}
\affil[1]{Institute of Biomedical Ethics and History of Medicine, University of Z\"urich, Z\"urich, Switzerland}
\affil[2]{SUPSI, Dalle Molle Institute for Artificial Intelligence (IDSIA), Lugano, Switzerland}
\affil[3]{ETH Z\"urich, Z\"urich, Switzerland}

\begin{document}

\maketitle

\begin{abstract}
 AI governance increasingly requires judgments about whether an AI system remains adequately trustworthy over time, whether observed changes are tolerable, and how such judgments should be documented in a transparent and contestable way. Existing approaches remain either too high-level to support lifecycle monitoring and reassessment or too narrowly metric-driven to connect multidimensional trustworthiness evidence with governance decisions. We propose a lightweight methodology centered on \emph{trustworthiness level functions}: auditable rules that map measured trustworthiness profiles to governance-relevant levels.  The methodology separates the underlying trustworthiness evidence from the governance rule used to interpret it and treats that rule as a lifecycle governance object. The rule may remain expert-defined or, when available evidence warrants empirical learning, be approximated by an interpretable candidate model. An AI lifecycle governance procedure embeds this choice in explicit decision gates for determining whether learning should be attempted and whether a learned candidate should become operative. The resulting rule supports lifecycle monitoring through level transitions, boundary margins, and profile drift, with explicit human responsibilities for validation, approval, and reassessment. We illustrate the methodology on synthetic AI lifecycle scenarios involving degradation, shocks, updates, heterogeneous monitoring cadences, and system comparison. Our methodology does not replace expert or legal judgment, but makes the governance interpretation of trustworthiness evidence more explicit, auditable, and contestable over time.
\end{abstract}


\section{Introduction}
\label{section:intro}
AI governance depends on judgments that are neither purely technical nor purely legal. Providers and oversight bodies must assess whether an AI system remains trustworthy after reconfiguration, updating, or shifts in its context of use. Here, we use AI trustworthiness in a governance-oriented sense: an AI system is trustworthy to the extent that, for a given intended purpose and context of use, it satisfies the socio-technical requirements documented in its design and governance. They must also determine when differences in trustworthiness across copies or variants of an AI system are material enough to warrant documentation, review, or oversight. Such trustworthiness-based questions arise across a wide range of systems, from LLM-based assistants and clinical decision-support systems to recommender agents. They are especially salient in settings where conformity assessment, monitoring, and reassessment become key AI lifecycle governance procedures, as in the case of high-risk AI systems under the EU AI Act (AIA) \citep{EU_AI_Act_2024}.

A central difficulty is that current work on AI trustworthiness is pulled in two directions. On the one hand, governance frameworks articulate trustworthiness through broad requirements such as robustness, transparency, fairness, and safety \citep{kemmerzell2025towards,mehrotra2026understanding}. On the other hand, technical work in system engineering and MLOps operationalizes subsets of these requirements through a plethora of local metrics, often without an explicit account of how such measurements should be integrated into governance over time \citep{rabanser2026towards,kreuzberger2023machine,bayram2024adaptive}. As a result, trustworthiness is frequently invoked in AI governance but rarely formalized in a way that supports simple and auditable AI lifecycle monitoring. Existing governance-relevant trustworthiness judgments focus on design or review activities \citep{kim2026toward}. However, even where individual---and possibly dozens of---trustworthiness dimensions can be measured over time \citep{kemmerzell2025towards,kaur2022trustworthy,mattioli2024overview}, their values do not by themselves determine whether the system remains within an accepted governance regime, has moved into a state requiring investigation, or has crossed a boundary warranting  reassessment. AI lifecycle governance therefore requires measurable trustworthiness dimensions \emph{and} explicit rules for interpreting their configurations in governance-relevant terms.

%%%%%%%%%%%%%%%%%%%%%%%%%%%%
% GAP
This paper addresses this gap by proposing \textbf{a  methodology for auditable trustworthiness levels in AI governance}. Its central object is a \emph{trustworthiness level function}: an explicit rule that maps a measured multidimensional trustworthiness profile to a governance-relevant trustworthiness level \citep{ferrario2025trustworthiness,ferrario2026highriskidentity}. The profile retains the underlying evidence about the system, whereas the level function specifies how that evidence is interpreted under a documented governance standard. Trustworthiness levels can therefore support lifecycle governance by making it possible to identify and audit regime membership, level transitions, proximity to governance boundaries, and changes that may warrant review or reassessment. However, we do not assume that the trustworthiness level function \emph{must} be learned from data. Where an expert-defined rule is already sufficiently explicit and operational, it can remain operative. Where expert judgment instead provides only coarse levels, anchor regions, constraints, or selected labeled profiles, empirical learning may be used to construct a candidate trustworthiness level function.
We embed this formalization in an \emph{AI lifecycle governance procedure} with explicit decision gates for determining whether empirical learning is warranted and whether a learned rule should replace the expert-defined one. The resulting trustworthiness level function supports post-deployment monitoring, documentation, and reassessment, providing an auditable layer between trustworthiness measurements and lifecycle governance decisions, including those relevant to conformity assessment and potential \emph{substantial modifications}---see Art.~3(23) and Art.~43(4) in the AIA and \citep{ferrario2026highriskidentity}. The paper makes the following contributions:

\begin{enumerate}

\item \textbf{Trustworthiness level functions as lifecycle governance objects.}
We separate multidimensional trustworthiness evidence from the rule used to interpret it and formalize this rule as an explicit, versioned, and auditable object for lifecycle governance.

\item \textbf{Conditional empirical learning of trustworthiness level functions.}
We characterize how candidate trustworthiness level functions may be learned from expert-labeled profiles and specify governance conditions for deciding whether learning is warranted and whether a learned candidate should be adopted.

\item \textbf{Lifecycle auditing and governance procedure.}
We embed expert-defined and empirically learned trustworthiness level functions in an AI lifecycle procedure for monitoring, documentation, validation, and reassessment, with explicit decision gates, governance controls, human responsibilities, and auditing measures.

\item \textbf{Illustrative lifecycle scenarios.}
We use synthetic AI lifecycle scenarios to illustrate cases in which empirical learning is informative, insufficient, or unnecessary, and to show how the proposed governance objects behave under degradation, shocks, updates, heterogeneous monitoring cadences, and different deployment contexts.
\end{enumerate}

%%%%%%%%%%%%%%%%%%%%%%%%%%%%%%%%%%%%%%
%%%%%%%%%%%%%%%%%%%%%%%%%%%%%%%%%%%%%%
\section{Background}
\label{section:background}

%%%%%%%%%%%%%%%%%%%%%%%%%%%%%%%%%%%%%%%
\subsection{AI Trustworthiness: Common Ground}
\label{subsection:TW_intro}

\emph{Trustworthy AI} is the paradigm that seeks to identify the criteria under which AI systems should be designed, developed, deployed, and governed so as to reflect core principles that ought to guide their role in society. 
% For instance, the AI HLEG framework is grounded in principles such as human agency and oversight, technical robustness and safety, privacy and data governance, transparency, diversity and fairness, societal and environmental well-being, and accountability \citep{EU_HLEG_TrustworthyAI_2019}. 
Over time, regulators, standardization bodies, governments, and international organizations have proposed partially overlapping approaches to delineating the scope of AI trustworthiness \citep{EU_HLEG_TrustworthyAI_2019,EU_AI_Act_2024,WhiteHouse2022AIBillOfRights,UNESCO2022EthicsAI,OECD2022AIRecommendation,ISOIEC24028_2020}. The literature on Trustworthy AI is now extensive: we refer to recent reviews for a deep dive into definitions and perspectives \citep{kaur2022trustworthy,kemmerzell2025towards,mehrotra2026understanding}. A common underlying assumption is that what makes an AI system worthy of trust is not a single property, but the satisfaction of \emph{multiple socio-technical requirements} \citep{mehrotra2026understanding,kemmerzell2025towards,kaur2022trustworthy}. The exact formulation of these requirements varies across frameworks, depends on how abstract principles are translated into operational criteria, and remains context-sensitive across domains. For instance, the  trustworthiness profile of an LLM-based self-study assistant will differ from that of a clinical decision-support system, or a recommender agent. That said, commonly discussed high-level AI trustworthiness dimensions include fairness, transparency, privacy and data governance, technical robustness and safety, accountability, human agency and oversight, and societal or environmental well-being \citep{kemmerzell2025towards,kaur2022trustworthy}. AI trustworthiness depends on the choice of individual dimensions and on how they interact, trade off, and reinforce one another over time.

A central implication is that trustworthiness must be embedded in an AI lifecycle governance capable of translating high-level principles into auditable practices \citep{mokander2022conformity,laux2024trustworthy,floridi2022capai}. The lifecycle perspective is important because AI systems are subject to change: conformity assessment establishes an initial baseline, post-deployment monitoring tracks continued compliance, and `substantial modification' can function as a governance-relevant discontinuity condition. This orientation is particularly visible in the EU AI Act---see Annex~IV and Article~72 therein \citep{EU_AI_Act_2024}. Thus, the challenge is to provide a sufficiently explicit representation of trustworthiness that can connect these obligations across time. 
% The next section assesses the extent to which this challenge is addressed by existing methods and metrics for evaluating AI trustworthiness.

%%%%%%%%%%%%%%%%%%%%%%%%%%%%%%%%%%%%%%%%%
\subsection{Existing Trustworthiness Assessment Tools, Procedures, and Audits}
\label{subsection:tw_tools}
Recent reviews suggest that the landscape of trustworthy-AI assessment is fragmented both instrumentally and conceptually \citep{kaur2022trustworthy,kemmerzell2025towards,mehrotra2026understanding}. In particular, Mehrotra et al's  recent scoping review of AIES and FAccT proceedings \citep{mehrotra2026understanding} shows that trustworthiness is a contested terrain where technical, ethical, and societal dimensions intersect.

Four families of approaches are relevant. First, there are governance-oriented checklists and assessment frameworks. In Europe, the Assessment List for Trustworthy AI (ALTAI) translates the AI HLEG requirements into a structured self-assessment instrument \citep{ALTAI2020,Radclyffe2023ALTAI}. In the United States, the NIST AI Risk Management Framework and its Playbook organize AI governance around the functions Govern, Map, Measure, and Manage \citep{NISTAIRMF2023}. 
These frameworks are valuable, but they remain primarily procedural: they do not themselves define explicit trustworthiness level rules, nor do they provide actionable diagnostics throughout the AI lifecycle. Second, there are documentation artifacts such as model cards, datasheets for datasets, and ethical codes \citep{Mitchell2019ModelCards,Gebru2021Datasheets,loi2019towards}. These improve transparency and accountability, yet their primary function is documentary rather than rule-inducing, e.g., by providing a dynamic rule for repeated trustworthiness assignment during post-deployment monitoring. Related governance instruments also use scoring and tiering to compress information into decision-relevant categories. For instance, the Canadian Algorithmic Impact Assessment is a questionnaire aggregating weighted information about an automated decision system into four impact levels that determine proportionate governance requirements \citep{CanadaAIA2026}, while the OECD Framework for the Classification of AI Systems organizes policy-relevant system characteristics across multiple dimensions to support risk assessment and regulatory differentiation \citep{OECD2022AIClassification}. NIST AI RMF Profiles similarly represent context-specific current or target states of AI risk-management practice \citep{NISTAIRMF2023}. Such approaches belong to a broader tradition of scorecards, risk matrices, and risk-tiering instruments that trade informational detail for more actionable governance categories \citep{cox2008risk}. In the algorithmic-accountability literature, however, this compression has also raised concerns that administratively tractable categories may obscure the judgments through which impacts and harms are constructed \citep{metcalf2021algorithmic}. However, recent FAccT work on index-based risk assessment shows that reasonable choices in constructing low-complexity multidimensional scores can materially alter resulting assessments and downstream decisions \citep{gosciak2026scrutinizing}. This makes the rule through which multidimensional evidence is translated into governance-relevant categories itself an object requiring explicit documentation and scrutiny.
Third, there are toolkits targeting specific trustworthiness dimensions, such as \texttt{AI Fairness 360}, \texttt{Fairlearn}, and \texttt{AI Explainability 360} \citep{Bellamy2019AIF360,Bird2020Fairlearn,Arya2019AIX360}. These are useful, but they remain dimension-specific and do not yield a unified, governance-level representation of system state  and its monitoring over time. Fourth, there are participatory and audit-style methodologies such as \texttt{Z-Inspection}, which emphasize interdisciplinary review and contextual evaluation \citep{zicari2021z}. Similarly, recent FAccT work has strengthened lifecycle-oriented AI assurance by specifying responsibilities, evidence collection, and structured assessment procedures across system development and deployment \citep{kim2026toward}. 
These approaches are valuable for surfacing ethical tensions, governing the assessment process---including expert judgment---across AI lifecycle stages. However, they do not yield a \emph{compact formal representation of trustworthiness}, nor rules for its monitoring over time.\footnote{Existing work on AI agents typically addresses only one trustworthiness layer---e.g., reliability metrics or runtime governance---without yielding a unified trustworthiness formalism for lifecycle monitoring \citep{rabanser2026towards,phiri2025creating,Staufer2026AIAgentIndex}.}

Taken together, four elements are rarely provided jointly. First, existing approaches only partially specify how heterogeneous trustworthiness dimensions should be integrated into one governance-relevant representation of system state. Second, even where such dimensions are measurable, they rarely provide an explicit and auditable rule for interpreting their joint configuration as governance-relevant levels such as acceptable, concerning, or risky. Third, post-deployment measurements may follow different evidential logics, delays, and monitoring cadences that require appropriate trustworthiness monitoring. Fourth, existing approaches only partially connect pre-deployment expert judgment, post-deployment monitoring, validation, rule revision, and reassessment within one documented lifecycle workflow.  Our methodology for auditable trustworthiness levels in AI lifecycle governance addresses these limitations.

%%%%%%%%%%%%%%%%%%%%%%%%%%%%%%%%%%%%%%%%
%%%%%%%%%%%%%%%%%%%%%%%%%%%%%%%%%%%%%%%%
\section{Formalization of AI Trustworthiness Levels}
\label{section:formalization}
In response to the four gaps identified above, we now define the formal objects used throughout the methodology. The aim is to define a lifecycle-ready representation of AI trustworthiness, an auditable rule for assigning trustworthiness levels, and a governance procedure that connects design-time labeling with post-deployment monitoring and reassessment.

%%%%%%%%%%%%%%%%%%%%%%%%%%%%%%
\subsection{AI Trustworthiness Commitments}
\label{subsection:AI_TW_Commitments}
Informed by the literature on Trustworthy AI, our methodology rests on six commitments. First, trustworthiness is context-sensitive and normative: relevant requirements depend on the intended purpose, application domain, and context of use of the AI system \citep{NISTAIRMF2023}. Second, if trustworthiness is to inform governance, it must be operationalized through indicators that can be assessed in practice \citep{ALTAI2020}. Third, trustworthiness is multidimensional and must therefore account for several requirements jointly rather than considering each dimension in isolation \citep{kaur2022trustworthy,ferrario2025trustworthiness}. A formal account therefore requires an explicit aggregation or regime-assignment rule. Fourth, perceived trustworthiness should be distinguished from assessed trustworthiness relative to a documented governance standard; here we focus on the latter \citep{liao2022designing,mehrotra2024systematic,ferrario2024justifying}.\footnote{Thus, here we focus on \emph{objective} trustworthiness, understood as the extent to which the system satisfies the trustworthiness-relevant requirements that hold in its context of use.} Fifth, trustworthiness is lifecycle-sensitive because AI systems and their deployment environments change over time \citep{kreuzberger2023machine,ferrario2024justifying}. Sixth, trustworthiness formalization should be governance-oriented: the point of formalization is to support documentation, assessment, monitoring, and reassessment in auditable workflows \citep{floridi2022capai,mokander2021ethics}.

%%%%%%%%%%%%%%%%%%%%%%%%%%%%%%%%%%%%%%%%%%%
\subsection{Protocols, Profiles, and Levels}
\label{subsection:protocol_profile_levels}
Let \(A\) be an AI system\footnote{An AI system is a machine-based system operating with varying degrees of autonomy, potentially adapting after deployment, and inferring from its inputs how to generate outputs, e.g., predictions, content, or recommendations, that may affect physical or virtual environments (Art. 3, \citep{EU_AI_Act_2024}).} with intended purpose \(I_A\) and context of use \(C_A\). We use \emph{intended purpose} in the governance-oriented sense familiar from the AIA: the use for which the AI system is intended by the provider, including the specific context and conditions of use as documented by the provider \citep{EU_AI_Act_2024}. We use \emph{context of use} more broadly to denote the realized socio-technical environment in which the system is deployed, monitored, and governed over time. Let
\(P_A=\{d_1,\dots,d_n\}\)
be $A$'s \emph{trustworthiness protocol}: a finite set of dimensions selected as relevant to the governance-relative trustworthiness of \(A\) in that context and their measurement protocols. 
% In this paper, this trustworthiness is understood in a governance-relative sense: it denotes the extent to which an AI system satisfies the trustworthiness-relevant requirements fixed by a documented protocol for a given intended purpose and context of use. (It is therefore objective relative to a governance standard.) 
Dimensions may include, for instance, robustness, fairness, transparency, or safety, but the protocol is not assumed universal as it is selected by context-sensitive criteria such as normative and operational relevance and auditability. 
For each dimension \(d_i\in P_A\), let \(k(i)\in\mathbb N\) denote the number of sub-metrics used to operationalize it.
Each sub-metric is normalized into \([0,1]\), giving a dimension vector
\(\mathbf q_i^A(t)=\bigl(q_{i,1}^A(t),\dots,q_{i,k(i)}^A(t)\bigr)\in[0,1]^{k(i)}.\)
Here, the normalization to \([0,1]\) is a convention; the formalization can equivalently be formulated on any common bounded scale, with thresholds and aggregation rules rescaled accordingly.

Furthermore, not all trustworthiness dimensions are measured in the same way. Some measures, such as explanation complexity or override rate (used to measure human agency and oversight), may be computed directly from model outputs, interface traces, or workflow logs available at time \(t\). Other measures, such as predictive accuracy, calibration, or outcome-based fairness measures, require realized outcomes and may therefore only be measurable over a recent window of resolved cases. Accordingly, \(\mathbf q_i^A(t)\) should be read as the most recent auditable estimate of dimension \(d_i\) available at time \(t\) under the measurement protocol adopted for that dimension, rather than as an instantaneous quantity computed from information available exactly at \(t\): Table~\ref{table:Q_dimensions}  in Appendix~\ref{appendix:table_TW} gives representative examples;  we refer to \citet{kemmerzell2025towards} for an extensive survey instead. Then, a dimension-level aggregator $a_i:[0,1]^{k(i)}\to[0,1]$ yields the dimension score
$Q_i^A(t):=a_i(\mathbf q_i^A(t))$ with the resulting dimension scores oriented so that higher values indicate greater trustworthiness.\footnote{Raw submetrics may retain their natural orientation. For instance, lower error rates, fairness gaps, or attack-success rates indicate better outcomes. Before entering the trustworthiness profile, such quantities are re-oriented at the dimension-processing stage so that higher dimension scores uniformly represent greater trustworthiness; for a normalized lower-is-better quantity 
\(q\in[0,1]\), this can be achieved by the transformation \(Q=1-q\).} Also, these aggregation rules depend on the dimension under consideration. For instance, when aggregating consistency or robustness measures for LLM-based applications, one may use averaging \citep{rabanser2026towards}. Aggregating safety-related measures may instead require the product of risk and expected severity measures \citep{kaplan1981quantitative}. Collecting these scores gives the observed \emph{AI trustworthiness profile}\footnote{At the theoretical level, we can represent the operative governance standard as a reference trustworthiness level rule $\mathsf{TW}^\ast:[0,1]^n \to [K]$, where \([K]=\{1,\dots,K\}\) is a fixed trustworthiness level scale chosen ex ante as a governance parameter. (Here, objectivity is with respect to a governance standard.) Along the trajectory of the system, $\mathsf{TW}^\ast(\mathbf Q_A(t))$ denotes the reference trustworthiness level of the system $A$ at time $t$ \citep{ferrario2024justifying,ferrario2025trustworthiness}.}
\(\mathbf Q_A(t):=\bigl(Q_1^A(t),\dots,Q_n^A(t)\bigr)\in[0,1]^n.\)

%%%%%%%%%%%%%%%%%%%%%%%%%%%%%%%%%%%%%%
%%%%%%%%%%%%%%%%%%%%%%%%%%%%%%%%%%%%%%
\subsection{Trustworthiness Level Functions as Governance Tools}
\label{subsection:tw_level_functions}
A trustworthiness profile represents multidimensional evidence about an AI system, but it does not by itself determine how that evidence should be interpreted for governance. With multiple measures per dimensions and possibly dozens of dimensions \citep{kemmerzell2025towards,kaur2022trustworthy,mattioli2024overview}, AI governance requires instruments to manage this complexity over time. Following \citep{ferrario2025trustworthiness}, we therefore introduce a \emph{trustworthiness level function}
\[
\mathsf{TW}:[0,1]^n\rightarrow[K],
\]
where the elements of \([K]:=\{1,\dots,K\}\) are ordered from lower to higher trustworthiness. The trustworthiness level of an AI system \(A\) at time \(t\) is \(\tau_A(t):=\mathsf{TW}\bigl(\mathbf Q_A(t)\bigr).\) While \citet{ferrario2025trustworthiness} uses \(\mathsf{TW}\) as a tool to define AI identity formally, here it becomes a tool for AI governance. The link between the two levels and its mapping into the EU AI Act pre- and post-deployment requirements for high-risk AI systems are provided in \citep{ferrario2026highriskidentity}. \(\mathsf{TW}\) partitions the trustworthiness profile space into level regions \(C_k:=\mathsf{TW}^{-1}(\{k\}),\ k\in[K].\) These regions make governance boundaries explicit and support the lifecycle diagnostics introduced below. The finite ordered codomain of \(\mathsf{TW}\) in \([K]\) is motivated by governance transparency. Similar to risk bucketing, a small number of discrete levels provides an interpretable and actionable representation of system state that can support monitoring, escalation, review, and reassessment. Therefore, while the underlying trustworthiness profile remains available for diagnosis and contestation, \(\mathsf{TW}\) specifies the governance rule through which that evidence is translated into a level.\footnote{We do not require every level in \([K]\) to occur in every deployment or lifecycle trajectory.} Conceptually, \(\mathsf{TW}\) is a \emph{governance regime assignment rule}, not an additional trustworthiness metric. The protocol \(P_A\) specifies what is measured, the profile \(\mathbf Q_A(t)\) records the resulting multidimensional evidence available at time \(t\), and \(\mathsf{TW}\) specifies how that evidence is interpreted as an ordered governance state. The three objects are therefore distinct: the same measured profile can in principle be interpreted under different documented governance standards, while a given trustworthiness level function can be repeatedly applied as new profiles become available over the system lifecycle. We impose one further minimal structural requirement. Assuming dimension scores are oriented so that higher values indicate better trustworthiness, \(\mathsf{TW}\) should be \emph{coordinatewise monotone}. For \(\mathbf q,\mathbf q'\in[0,1]^n\), let
\(\mathbf q\preceq\mathbf q'\Longleftrightarrow q_i\leq q_i'
\ \text{for all }i.\) We require \(\mathbf q\preceq\mathbf q'\Longrightarrow \mathsf{TW}(\mathbf q)\leq\mathsf{TW}(\mathbf q'). \)
Thus, improving one or more dimensions while holding the others constant cannot lower the assigned trustworthiness level: the system either remains in the same regime or moves to a higher one. However, coordinatewise monotonicity does not order profiles that trade improvements in some dimensions against deterioration in others. Therefore, in such cases, the explicit assignments made by \(\mathsf{TW}\) encode additional governance judgments that must be made explicit in its construction. We tackle this construction in what follows.

%%%%%%%%%%%%%%%%%%%%%%%%%%%%%%%%%%%%%%%%%%%%%%%%%%%
% FROM PROTOCOL TO TIME SERIES TO TRUSTWORTHINESS LEVEL FUNCTION
%%%%%%%%%%%%%%%%%%%%%%%%%%%%%%%%%%%%%%%%%%%%%%%%%%%

% View angles for the cube (panel c)
\tdplotsetmaincoords{70}{45}

\begin{figure*}[t]
\centering
\tikzset{
  paneltitle/.style={font=\small\bfseries},
  lab/.style={font=\scriptsize},
  axis3d/.style={-{Stealth[length=2.1mm]}, line width=0.7pt},
  axis3dsoft/.style={-{Stealth[length=1.6mm]}, line width=0.35pt, draw=black!45},
  cubeface/.style={draw=black, line width=0.4pt},
  regedge/.style={draw=black, line width=0.4pt},
  levellab/.style={font=\scriptsize\bfseries},
  traj/.style={densely dotted, line width=0.55pt},
  protocolbox/.style={draw, rounded corners=2mm, line width=0.5pt, inner sep=5pt}
}

%%%%%%%%%%%%%%%%%%%%%%%%%%%%%%%%%%%%%%%%%%%%%%%%
% PANEL LABELS
%%%%%%%%%%%%%%%%%%%%%%%%%%%%%%%%%%%%%%%%%%%%%%%%
\makebox[\textwidth][c]{%
\begin{minipage}{0.20\textwidth}
\centering
\small\bfseries (a)
\end{minipage}%
\begin{minipage}{0.30\textwidth}
\centering
\small\bfseries (b)
\end{minipage}%
\begin{minipage}{0.40\textwidth}
\centering
\small\bfseries (c)
\end{minipage}%
}

\vspace{-0.35em}

\makebox[\textwidth][c]{%

%%%%%%%%%%%%%%%%%%%%%%%%%%%%%%%%%%%%%%%%%%%%%%%%
% PANEL (a): PROTOCOL P
%%%%%%%%%%%%%%%%%%%%%%%%%%%%%%%%%%%%%%%%%%%%%%%%
\begin{minipage}[c]{0.20\textwidth}
\centering
\begin{tikzpicture}
\node[
  protocolbox,
  align=left,
  minimum width=0.88\linewidth,
  font=\scriptsize
] (Pbox) {%
{\bfseries Protocol \(P\)}\\[0.35em]
\textbf{Accuracy} \((q_{1,1},\,q_{1,2})\)\\[0.25em]
\textbf{Robustness} \((q_{2,1},\,q_{2,2})\)\\[0.25em]
\textbf{Fairness} \((q_{3,1},\,q_{3,2})\)\\[0.25em]
\hspace*{0.9cm}\(\cdots\)
};
\end{tikzpicture}
\end{minipage}%

%%%%%%%%%%%%%%%%%%%%%%%%%%%%%%%%%%%%%%%%%%%%%%%%
% PANEL (b): THREE TIME SERIES
%%%%%%%%%%%%%%%%%%%%%%%%%%%%%%%%%%%%%%%%%%%%%%%%
\begin{minipage}[c]{0.30\textwidth}
\centering
\begin{tikzpicture}
\begin{groupplot}[
  group style={group size=1 by 3, vertical sep=0.60cm},
  width=0.95\linewidth,
  height=2.65cm,
  xmin=0, xmax=10,
  ymin=0, ymax=1,
  axis lines=left,
  enlargelimits=false,
  tick align=outside,
  xtick={0,2,4,6,8,10},
  ytick={0,1},
  tick label style={font=\scriptsize},
  label style={font=\scriptsize},
  every axis y label/.style={at={(-0.12,0.5)},font=\tiny},
  every axis x label/.style={
    at={(axis description cs:1.02,-0.08)},
    anchor=west,
    font=\scriptsize
  }
]

% --- Accuracy
\nextgroupplot[
  ylabel={$Q_1(t)$},
  xticklabels=\empty,
  title={\scriptsize Accuracy}
]
\addplot[line width=0.9pt, smooth] coordinates {
  (0,0.93) (1,0.92) (2,0.92) (3,0.90) (4,0.86)
  (5,0.78) (6,0.74) (7,0.79) (8,0.82) (9,0.79) (10,0.76)
};
\addplot[only marks, mark=*, mark size=1.2pt] coordinates {
  (0.5,0.93) (1.5,0.92) (2.5,0.91) (3.5,0.88) (4.5,0.82)
  (5.5,0.76) (6.5,0.77) (7.5,0.81) (8.5,0.80) (9.5,0.77)
};

% --- Robustness
\nextgroupplot[
  ylabel={$Q_2(t)$},
  xticklabels=\empty,
  title={\scriptsize Robustness}
]
\addplot[line width=0.9pt, smooth] coordinates {
  (0,0.89) (1,0.89) (2,0.88) (3,0.86) (4,0.82)
  (5,0.76) (6,0.71) (7,0.74) (8,0.77) (9,0.74) (10,0.72)
};
\addplot[only marks, mark=*, mark size=1.2pt] coordinates {
  (1.0,0.89) (3.0,0.86) (5.0,0.76) (7.0,0.74) (9.0,0.74)
};

% --- Fairness
\nextgroupplot[
  ylabel={$Q_3(t)$},
  xlabel={time},
  title={\scriptsize Fairness}
]
\addplot[line width=0.9pt, smooth] coordinates {
  (0,0.87) (1,0.87) (2,0.86) (3,0.85) (4,0.81)
  (5,0.76) (6,0.72) (7,0.75) (8,0.77) (9,0.75) (10,0.74)
};
\addplot[only marks, mark=*, mark size=1.2pt] coordinates {
  (0.8,0.87) (2.2,0.86) (3.7,0.84) (5.2,0.75)
  (6.7,0.73) (8.3,0.77) (9.6,0.74)
};

\end{groupplot}
\end{tikzpicture}
\end{minipage}%

%%%%%%%%%%%%%%%%%%%%%%%%%%%%%%%%%%%%%%%%%%%%%%%%
% PANEL (c): 3D CUBE WITH FOUR LEVELS + TRAJECTORY
%%%%%%%%%%%%%%%%%%%%%%%%%%%%%%%%%%%%%%%%%%%%%%%%
\begin{minipage}[c]{0.40\textwidth}
\centering
\begin{tikzpicture}[
  tdplot_main_coords,
  scale=2.80,
  line join=round,
  line cap=round
]

% Thresholds
\def\xcut{0.60}
\def\ycut{0.60}
\def\zcut{0.70}

% Trustworthiness-level function
\node[
  font=\scriptsize,
  align=center
] at (0.55,0.60,1.35)
{\(\mathsf{TW}:[0,1]^3\to[4]\), \quad
 \(C_k=\mathsf{TW}^{-1}(\{k\})\)};

% ------------------------------------------------
% FILLED REGION FACES
% Regions:
% C1: x < xcut
% C2: x >= xcut, y < ycut
% C3: x >= xcut, y >= ycut, z < zcut
% C4: x >= xcut, y >= ycut, z >= zcut
% ------------------------------------------------

% ---------- C1 ----------
\filldraw[cubeface,fill=black!24,opacity=0.76]
  (0,0,0) -- (0,1,0) -- (0,1,1) -- (0,0,1) -- cycle;
\filldraw[cubeface,fill=black!24,opacity=0.76]
  (\xcut,0,0) -- (\xcut,1,0) -- (\xcut,1,1) -- (\xcut,0,1) -- cycle;
\filldraw[cubeface,fill=black!24,opacity=0.76]
  (0,0,0) -- (\xcut,0,0) -- (\xcut,0,1) -- (0,0,1) -- cycle;
\filldraw[cubeface,fill=black!24,opacity=0.76]
  (0,1,0) -- (\xcut,1,0) -- (\xcut,1,1) -- (0,1,1) -- cycle;
\filldraw[cubeface,fill=black!24,opacity=0.76]
  (0,0,0) -- (\xcut,0,0) -- (\xcut,1,0) -- (0,1,0) -- cycle;
\filldraw[cubeface,fill=black!24,opacity=0.76]
  (0,0,1) -- (\xcut,0,1) -- (\xcut,1,1) -- (0,1,1) -- cycle;

% ---------- C2 ----------
\filldraw[cubeface,fill=black!16,opacity=0.79]
  (\xcut,0,0) -- (\xcut,\ycut,0) -- (\xcut,\ycut,1) -- (\xcut,0,1) -- cycle;
\filldraw[cubeface,fill=black!16,opacity=0.79]
  (1,0,0) -- (1,\ycut,0) -- (1,\ycut,1) -- (1,0,1) -- cycle;
\filldraw[cubeface,fill=black!16,opacity=0.79]
  (\xcut,0,0) -- (1,0,0) -- (1,0,1) -- (\xcut,0,1) -- cycle;
\filldraw[cubeface,fill=black!16,opacity=0.79]
  (\xcut,\ycut,0) -- (1,\ycut,0) -- (1,\ycut,1) -- (\xcut,\ycut,1) -- cycle;
\filldraw[cubeface,fill=black!16,opacity=0.79]
  (\xcut,0,0) -- (1,0,0) -- (1,\ycut,0) -- (\xcut,\ycut,0) -- cycle;
\filldraw[cubeface,fill=black!16,opacity=0.79]
  (\xcut,0,1) -- (1,0,1) -- (1,\ycut,1) -- (\xcut,\ycut,1) -- cycle;

% ---------- C3 ----------
\filldraw[cubeface,fill=black!10,opacity=0.82]
  (\xcut,\ycut,0) -- (\xcut,1,0) -- (\xcut,1,\zcut) -- (\xcut,\ycut,\zcut) -- cycle;
\filldraw[cubeface,fill=black!10,opacity=0.82]
  (1,\ycut,0) -- (1,1,0) -- (1,1,\zcut) -- (1,\ycut,\zcut) -- cycle;
\filldraw[cubeface,fill=black!10,opacity=0.82]
  (\xcut,\ycut,0) -- (1,\ycut,0) -- (1,\ycut,\zcut) -- (\xcut,\ycut,\zcut) -- cycle;
\filldraw[cubeface,fill=black!10,opacity=0.82]
  (\xcut,1,0) -- (1,1,0) -- (1,1,\zcut) -- (\xcut,1,\zcut) -- cycle;
\filldraw[cubeface,fill=black!10,opacity=0.82]
  (\xcut,\ycut,0) -- (1,\ycut,0) -- (1,1,0) -- (\xcut,1,0) -- cycle;
\filldraw[cubeface,fill=black!10,opacity=0.82]
  (\xcut,\ycut,\zcut) -- (1,\ycut,\zcut) -- (1,1,\zcut) -- (\xcut,1,\zcut) -- cycle;

% ---------- C4 ----------
\filldraw[cubeface,fill=diagramblue!18,opacity=0.72]
  (\xcut,\ycut,\zcut) -- (\xcut,1,\zcut) -- (\xcut,1,1) -- (\xcut,\ycut,1) -- cycle;
\filldraw[cubeface,fill=diagramblue!18,opacity=0.72]
  (1,\ycut,\zcut) -- (1,1,\zcut) -- (1,1,1) -- (1,\ycut,1) -- cycle;
\filldraw[cubeface,fill=diagramblue!18,opacity=0.72]
  (\xcut,\ycut,\zcut) -- (1,\ycut,\zcut) -- (1,\ycut,1) -- (\xcut,\ycut,1) -- cycle;
\filldraw[cubeface,fill=diagramblue!18,opacity=0.72]
  (\xcut,1,\zcut) -- (1,1,\zcut) -- (1,1,1) -- (\xcut,1,1) -- cycle;
\filldraw[cubeface,fill=diagramblue!18,opacity=0.72]
  (\xcut,\ycut,\zcut) -- (1,\ycut,\zcut) -- (1,1,\zcut) -- (\xcut,1,\zcut) -- cycle;
\filldraw[cubeface,fill=diagramblue!18,opacity=0.72]
  (\xcut,\ycut,1) -- (1,\ycut,1) -- (1,1,1) -- (\xcut,1,1) -- cycle;

% ------------------------------------------------
% REGION EDGES: ALL SOLID, SAME STYLE
% ------------------------------------------------

% Outer cube edges
\draw[regedge] (0,0,0) -- (1,0,0);
\draw[regedge] (1,0,0) -- (1,1,0);
\draw[regedge] (1,1,0) -- (0,1,0);
\draw[regedge] (0,1,0) -- (0,0,0);
\draw[regedge] (0,0,1) -- (1,0,1);
\draw[regedge] (1,0,1) -- (1,1,1);
\draw[regedge] (1,1,1) -- (0,1,1);
\draw[regedge] (0,1,1) -- (0,0,1);
\draw[regedge] (0,0,0) -- (0,0,1);
\draw[regedge] (1,0,0) -- (1,0,1);
\draw[regedge] (1,1,0) -- (1,1,1);
\draw[regedge] (0,1,0) -- (0,1,1);

% C1 edges
\draw[regedge] (\xcut,0,0) -- (\xcut,1,0);
\draw[regedge] (\xcut,1,0) -- (\xcut,1,1);
\draw[regedge] (\xcut,0,0) -- (\xcut,0,1);
\draw[regedge] (\xcut,0,1) -- (\xcut,1,1);

% C2 edges
\draw[regedge] (1,\ycut,0) -- (1,\ycut,1);
\draw[regedge] (\xcut,\ycut,0) -- (\xcut,\ycut,1);
\draw[regedge] (\xcut,\ycut,0) -- (1,\ycut,0);
\draw[regedge] (\xcut,\ycut,1) -- (1,\ycut,1);

% C3 / C4 boundary edges
\draw[regedge] (\xcut,\ycut,\zcut) -- (1,\ycut,\zcut);
\draw[regedge] (\xcut,1,\zcut) -- (1,1,\zcut);
\draw[regedge] (\xcut,\ycut,\zcut) -- (\xcut,1,\zcut);
\draw[regedge] (1,\ycut,\zcut) -- (1,1,\zcut);

% C4 vertical edges
\draw[regedge] (\xcut,\ycut,\zcut) -- (\xcut,\ycut,1);
\draw[regedge] (\xcut,1,\zcut) -- (\xcut,1,1);
\draw[regedge] (1,\ycut,\zcut) -- (1,\ycut,1);
\draw[regedge] (1,1,\zcut) -- (1,1,1);

% ------------------------------------------------
% Region labels
% ------------------------------------------------
\node[levellab] at (0.10,0.76,0.90) {$C_1$};
\node[levellab] at (0.60,0.22,0.50) {$C_2$};
\node[levellab] at (0.99,0.82,0.30) {$C_3$};
\node[levellab,text=diagramblue] at (0.82,0.87,0.98) {$C_4$};

% ------------------------------------------------
% Trajectory inside C4
% ------------------------------------------------
\coordinate (T1) at (0.93,0.89,0.87);
\coordinate (T2) at (0.89,0.84,0.84);
\coordinate (T3) at (0.81,0.77,0.78);
\coordinate (T4) at (0.74,0.71,0.72);
\coordinate (T5) at (0.80,0.76,0.76);
\coordinate (T6) at (0.76,0.72,0.74);

\draw[traj] (T1) -- (T2) -- (T3) -- (T4) -- (T5) -- (T6);

\fill[black] (T1) circle (0.55pt);
\fill[black] (T2) circle (0.55pt);
\fill[black] (T3) circle (0.55pt);
\fill[black] (T4) circle (0.55pt);
\fill[black] (T5) circle (0.55pt);
\fill[black] (T6) circle (0.55pt);

% Axes
\draw[axis3d] (0,0,0) -- (1.12,0,0)
  node[anchor=west] {$Q_1$};

\draw[axis3dsoft] (0,0,0) -- (0,1.12,0)
  node[anchor=south,font=\scriptsize,text=black!55] {$Q_2$};

\draw[axis3d] (0,0,0) -- (0,0,1.15)
  node[anchor=east] {$Q_3$};

\end{tikzpicture}
\end{minipage}%
}

%\vspace{-0.3em}

\caption{From trustworthiness protocol to trustworthiness level assignment.
(a) A trustworthiness protocol \(P\) specifies relevant dimensions together with the underlying measures used to operationalize them. (b) These dimensions generate time-dependent trustworthiness signals \(Q_i(t)\), observed through measurements collected at different cadences.(c) The resulting profile \(\mathbf Q(t)=(Q_1(t),Q_2(t),Q_3(t))\) is mapped by a trustworthiness level function \(\mathsf{TW}\) to one of four regions \(C_1,\dots,C_4\), where \(C_k=\mathsf{TW}^{-1}(\{k\})\). The dotted trajectory illustrates the evolution of one AI system over time within the highest trustworthiness region \(C_4\).}
\label{fig:tw_protocol_timeseries_function}
\end{figure*}
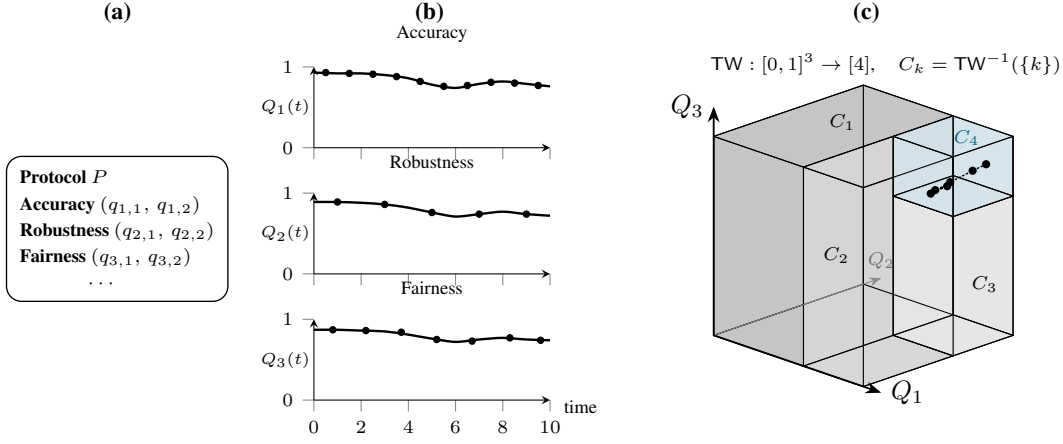
%%%%%%%%%%%%%%%%%%%%%%%%%%%%%%%%%%%%%%%%%%%%%%%%%%%

%%%%%%%%%%%%%%%%%%%%%%%%%%%%%%%%%%%%%%
\subsection{Expert-Governance versus Empirically-Learned  Trustworthiness Level Functions}
\label{subsection:learning_TW}
How should a trustworthiness level function be constructed in practice? We distinguish two possibilities.

\paragraph{Expert-driven trustworthiness level rules.}
When level assignment is determined through a documented expert governance procedure, we denote the resulting rule by \(\mathsf{TW}^{\mathrm{exp}}\). This rule need not be available as a closed-form mathematical expression: it may be instantiated through explicit thresholds, anchor regions,   constraints or a documented adjudication procedure. These expert-driven rules arise naturally in real-world governance settings. In medicine, assigning a trustworthiness level to a clinical decision-support system may require specialist chart review, adjudication of delayed patient outcomes, subgroup analysis, and joint assessment by clinicians, data scientists, safety officers, and ethics or compliance personnel. Comparable procedures arise in industrial or public-sector applications, where engineering, legal, operational, and stakeholder expertise may be required to determine whether a system state remains acceptable. The same general logic is already familiar in AI governance practice: before deployment, expert committees, red-teaming exercises, or internal review procedures often identify clearly acceptable, borderline, or unacceptable system behaviors without attempting to map the entire space of possible cases in a fully granular way \citep{zicari2021z,NISTAIRMF2023}. Expert rules are therefore often coarse rather than exhaustive. Pre-deployment evidence may cover only a limited part of the trustworthiness-profile space, and borderline cases may require further review. Nevertheless, these rules provide the documented and contestable governance basis from which trustworthiness levels can initially be assigned. Their labels and criteria may subsequently require revision as the system, its environment, evidence base, or governance standard change \citep{zicari2021z,NISTAIRMF2023}.

\paragraph{Empirically learned \(\widehat{\mathsf{TW}}\).}
Suppose that an expert governance procedure yields a historical labeled dataset \(\mathcal D_N=\bigl\{(\mathbf Q_A(t_i),Y_i)\bigr\}_{i=1}^N,\) where \(\mathbf Q_A(t_i)\in[0,1]^n\) is an observed trustworthiness profile and \(Y_i\in[K]\) is its expert-assigned trustworthiness level. The labels may be produced through an explicit expert rule \(\mathsf{TW}^{\mathrm{exp}}\) or through documented adjudication of observed profiles over time. Where empirical learning is warranted---in Section~\ref{subsection:AI_governance_phases} we will discuss this point extensively---the goal is to construct a candidate trustworthiness level function
\[
\widehat{\mathsf{TW}}:[0,1]^n\rightarrow[K]
\]
from an admissible class \(\mathcal H\) of functions. A generic learning problem is
\begin{equation}
\widehat{\mathsf{TW}}
\in
\operatorname*{arg\,min}_{h\in\mathcal H}
\left\{
\frac{1}{N}
\sum_{i=1}^{N}
\ell\bigl(h(\mathbf Q_A(t_i)),Y_i\bigr)
+
\lambda\Omega(h)
\right\},
\label{eq:problem}
\end{equation}
where \(\ell\) measures disagreement with the expert labels and \(\Omega(h)\) penalizes rule complexity. Because the levels are ordered, the loss may assign larger penalties to errors spanning more distant trustworthiness levels.\footnote{One can also consider penalties that reflect the special normative status of selected trustworthiness dimensions.} The learning procedure does not prescribe a unique model class \(\mathcal H\). In this work we consider shallow decision trees, see Section~\ref{section:empirical_study}, because threshold-based rules are transparent, easily documented, and naturally aligned with governance standards expressed through conditions on individual dimensions \citep{breiman2017classification}. However, decision trees should be abandoned when to-be-learned rules are too high-dimensional, interaction-heavy, or non-axis-aligned. In such settings, more complex models should be considered. In all cases, candidate functions should be evaluated for interpretability, temporal generalization, regime coverage, and stability under small admissible perturbations of profiles and training data. In particular, a good learned rule should not oscillate sharply between adjacent trustworthiness levels because one dimension crosses the boundary repeatedly under minor measurement variation or small sample perturbations. Candidate rules should therefore also be stress-tested for \emph{boundary robustness}: one should prefer rules whose level assignments remain stable under small admissible perturbations of trustworthiness profiles and under modest changes in the training data. Rules that exhibit short-horizon zig-zag transitions between levels under such perturbations should be treated as governance-inadequate and discarded \citep{ferrario2025trustworthiness}. Among comparably performing candidates, governance actors should prefer learned trustworthiness level functions whose regime transitions are sparse, interpretable, and robust.

\paragraph{Expert-driven or empirically learned?}
Empirical learning is neither required nor intrinsically preferable. If \(\mathsf{TW}^{\mathrm{exp}}\) is sufficiently explicit, operational, and auditable, retaining it may be the appropriate choice. Learning a function \(\widehat{\mathsf{TW}}\) becomes potentially useful when expert judgment is costly or sparse and the available profiles, labels, and regime variation support meaningful empirical generalization. Conversely, weak regime coverage, unstable labels, poorly measured dimensions, or insufficient validation evidence may make learning uninformative. Accordingly, \(\widehat{\mathsf{TW}}\) should be understood as a \emph{candidate} operationalization of an expert-driven governance standard whose adoption requires governance approval. In general, it may (1) reproduce that standard, (2) further compress distinctions weakly supported by the available evidence, or (3) suggest changes in the partition of trustworthiness space. The procedure in Section~\ref{section:governance_procedure} determines which rule should become operative in a given context.

%%%%%%%%%%%%%%%%%%%%%%%%%%%%%%%%%%%%%%%%
%%%%%%%%%%%%%%%%%%%%%%%%%%%%%%%%%%%%%%%%
\section{AI Lifecycle Governance Procedure}
\label{section:governance_procedure}
We embed our formalization within a \emph{lightweight AI lifecycle governance procedure}, showing how it provides a quantitative and auditable layer for lifecycle governance that can be aligned with conformity assessment,  post-deployment monitoring, and reassessment procedures across AI systems. This concludes the exposition of our methodology.

%%%%%%%%%%%%%%%%%%%%%%%%%%%%%%%
\subsection{Structure of the AI Lifecycle Governance Procedure}
\label{subsection:AI_governance_phases}
\begin{figure*}[t]
\centering

\begin{tikzpicture}
[scale=0.65,
transform shape,
>=Latex,
node distance=5mm and 6mm,
box/.style=
{draw,rounded corners=2mm,align=center,minimum height=8.5mm,text width=3.15cm,inner sep=3.5pt},
gate/.style={draw,diamond,aspect=2,align=center,inner sep=1.2pt,text width=2.35cm},
flow/.style={-{Latex[length=1.1mm,width=0.75mm]},
line width=0.28pt},
back/.style={draw,densely dotted,line width=0.30pt,rounded corners=3mm,inner sep=5pt}
]

% --- Top row: pre-deployment governance
\node[box,fill=blue!8] (sys) {Specify AI system\\intended purpose and context of use};

\node[box,fill=blue!8,right=of sys] (protocol) {Define trustworthiness protocol\\and reference sources};

\node[box,fill=blue!8,right=of protocol] (measure) {Operationalize dimensions,\\measures, and aggregation};

\node[box,fill=blue!8,right=of measure] (label) {Define level scale,\\auditable labeling procedure,\\and expert rule $\mathsf{TW}^{\mathrm{exp}}$};

\node[box,fill=blue!8,right=of label] (collect) {Collect labeled pre-deployment data\\(e.g., sandbox, shadow mode, pilot)};

% --- Middle row: conditional learning
\node[box,fill=orange!10,below=29mm of sys] (dataset) {Historical labeled dataset\\$\mathcal D_N=\{(\mathbf Q_A(t_i),Y_i)\}_{i=1}^N$};

\node[gate,fill=orange!10,right=of dataset] (gate1) {\textbf{Gate 1}\\Is empirical\\learning warranted?};

\node[box,fill=orange!10,right=of gate1] (learn) {Learn candidate rule\\$\widehat{\mathsf{TW}}:[0,1]^n\to[K]$};

\node[box,fill=orange!10,right=of learn] (validate) {Document model structure\\and validation performance};

\node[gate,fill=orange!10,right=of validate] (gate2) {\textbf{Gate 2}\\Adopt learned\\rule?};

% --- Bottom row: deployment and monitoring
\node[box,fill=green!10,below=65mm of sys] (choice) {Set operative rule:\\$\mathsf{TW}^{\mathrm{op}}\in\{\mathsf{TW}^{\mathrm{exp}},\widehat{\mathsf{TW}}\}$};

\node[box,fill=green!10,right=of choice] (deploy) {Deploy chosen\\governance rule};

\node[box,fill=green!10,right=of deploy] (monitor) {Measure profile over time\\(logs, windows, delayed labels)};

\node[box,fill=green!10,right=of monitor] (assess) {Assign levels, margins, and drift\\under the operative rule};

\node[box,fill=green!10,right=of assess] (report) {Report outputs and trigger\\review, validation, or dataset update};

% --- Flows top row
\draw[flow] (sys) -- (protocol);
\draw[flow] (protocol) -- (measure);
\draw[flow] (measure) -- (label);
\draw[flow] (label) -- (collect);

% --- Top to middle
\draw[flow,draw=black!55] (collect.south) -- ++(0,-10mm) -| (dataset.north);

% --- Middle row
\draw[flow] (dataset) -- (gate1);
\draw[flow] (gate1) -- node[above,font=\footnotesize]{\textbf{Yes}} (learn);
\draw[flow] (learn) -- (validate);
\draw[flow] (validate) -- (gate2);

% --- Gate 1: no learning -> expert rule remains operative
\draw[flow,draw=black!55] (gate1.south) -- ++(0,-12mm) -| node[pos=-0.07,font=\footnotesize]{\textbf{No}: expert} ([xshift=8mm]choice.north);

% --- Gate 2: learned rule accepted -> learned rule operative
\draw[flow,draw=black!55] (gate2.south) -- ++(0,-7mm) -| node[pos=0.0,right,font=\footnotesize]{\textbf{Yes}: learned} ([xshift=4mm]choice.north);

% --- Gate 2: learned rule rejected -> expert rule operative
\coordinate (gate2rejectlow) at ([yshift=-11mm]choice.south);

\draw[flow,draw=black!55]
  (gate2.east) -- ++(10mm,0)
  |- node[pos=0.48,left,font=\footnotesize]{\textbf{No}: expert}
  (gate2rejectlow)
  -- (choice.south);

% --- Bottom row
\draw[flow] (choice) -- (deploy);
\draw[flow] (deploy) -- (monitor);
\draw[flow] (monitor) -- (assess);
\draw[flow] (assess) -- (report);

% --- Monitoring feedback to dataset
\draw[-{Latex[length=1.1mm,width=0.75mm]},draw=black!55,line width=0.28pt] (report.north) -- ++(0,11mm) -| node[pos=0.15,above,align=right,name=datasetfeedback,xshift=2mm,yshift=-1mm] {\footnotesize reviewed cases may update dataset} (dataset.south);

% --- Feedback to labeling procedure
\draw[
  -{Latex[length=1.1mm,width=0.75mm]},
  draw=black!55,
  line width=0.28pt
]
(report.east)
-- ++(36mm,0)
|- node[pos=0.40,left,align=left]
   {\footnotesize update\\
    \footnotesize labeling\\
    \footnotesize procedure}
([yshift=8mm]label.north)
-- (label.north);

% --- Finely and densely dotted phase rectangles
\node[back,fit=(sys)(protocol)(measure)(label)(collect)] (govbox) {};

\node[back,fit=(dataset)(gate1)(learn)(validate)(gate2)] (learnbox) {};

\coordinate (depboxSW) at ([yshift=-1mm]choice.south west);

\coordinate (depboxSE) at ([yshift=-1mm]report.south east);
\node[back,fit=(choice)(deploy)(monitor)(assess)(report)(depboxSW)(depboxSE)] (depbox) {};

% --- Phase labels
\node[anchor=south west,align=left] at ([xshift=1mm,yshift=-1mm]govbox.north west) {\textbf{Phase 1: Pre-deployment governance and auditing procedure}};

\node[anchor=south west,align=left] at ([xshift=1mm,yshift=-1mm]learnbox.north west) {\textbf{Phase 2: Conditional empirical learning and rule selection}};

\node[anchor=south west,align=left] at ([xshift=2mm,yshift=-1mm]depbox.north west) {\textbf{Phase 3: Post-deployment monitoring and reassessment}};

% --- Global outer frame
\node[thick,rounded corners=2mm,inner sep=16pt,fit=(govbox)(learnbox)(depbox)(datasetfeedback)] (outerbox) {};

\draw[thick,rounded corners=2mm] (outerbox.south west) rectangle (outerbox.north east);

\node[anchor=south,align=center,font=\Large] at ([yshift=1mm]outerbox.north) {\textbf{AI Lifecycle Governance Procedure}};

\end{tikzpicture}
\caption{AI lifecycle governance procedure. \textbf{Phase 1} specifies the trustworthiness protocol, measures, level scale, and expert-defined rule $\mathsf{TW}^{\mathrm{exp}}$. \textbf{Phase 2} introduces two explicit decision gates: Gate~1 determines whether empirical learning is warranted; Gate~2 determines whether a learned candidate $\widehat{\mathsf{TW}}$ should be adopted. \textbf{Phase 3} monitors the operative rule $\mathsf{TW}^{\mathrm{op}}$, whether expert-defined or learned, and allows reviewed cases to trigger dataset, protocol, or reassessment updates.}
\label{fig:workflow_governance}
\end{figure*}
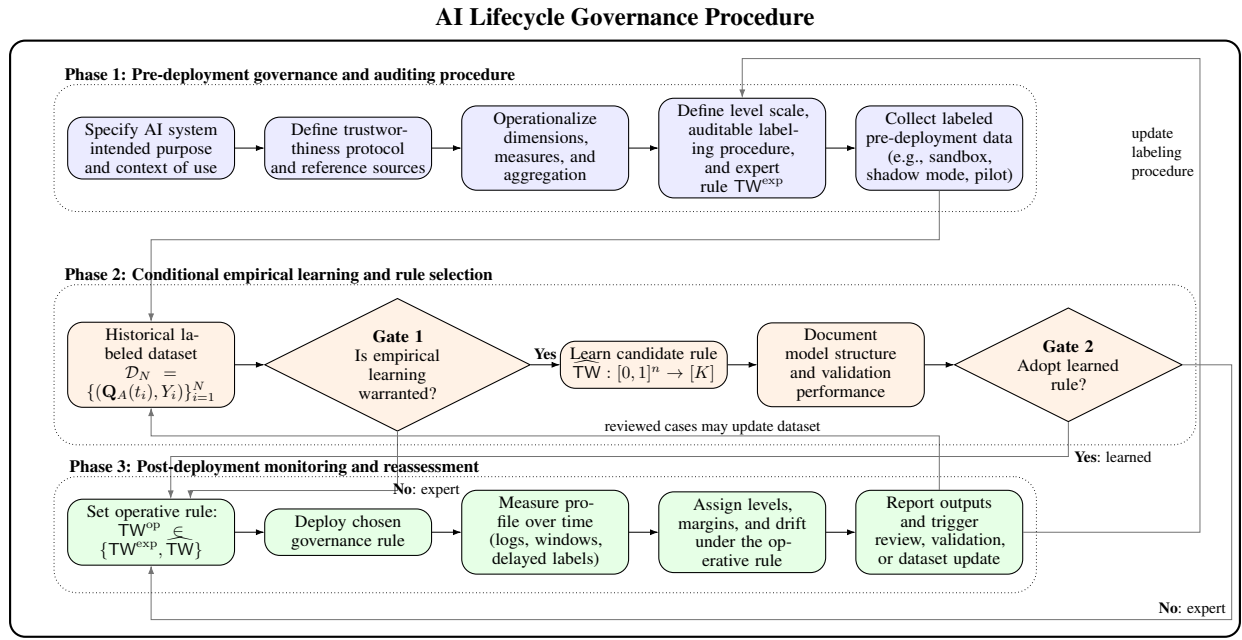

Figure~\ref{fig:workflow_governance} summarizes a lightweight AI lifecycle governance procedure in three phases and two decision gates. The procedure distinguishes: (1) the construction of an expert-defined trustworthiness standard \(\mathsf{TW}^{\mathrm{exp}}\), (2) the conditional learning of a candidate empirical function \(\widehat{\mathsf{TW}}\), and (3) the selection and monitoring of an operative rule
\[
\mathsf{TW}^{\mathrm{op}}
\in
\left\{
\mathsf{TW}^{\mathrm{exp}},
\widehat{\mathsf{TW}}
\right\}.
\]
The two gates determine whether empirical learning is warranted and, if so, whether the learned rule is sufficiently adequate for governance use. Our AI lifecycle governance procedure comprises three phases:

\paragraph{Phase 1: Pre-deployment governance and auditing procedure.}
This phase establishes the trustworthiness standard to be used for governance. Once the AI system, its intended purpose, and its context of use have been specified \citep{EU_AI_Act_2024}, governance actors define the trustworthiness protocol, operationalize its dimensions through measures and aggregation rules, and determine the trustworthiness level scale together with the labeling procedure discussed in Section~\ref{subsection:learning_TW}. The resulting expert-defined rule \(\mathsf{TW}^{\mathrm{exp}}\) provides the initial basis for assigning trustworthiness levels. In regulated settings, these design choices also support conformity documentation by making explicit which dimensions are monitored, how they are measured, and how level assignments are justified---see Annex~IV, Sections~1--4 in the AIA.

\paragraph{Phase 2: Conditional empirical learning and rule selection.}
In this phase, trustworthiness-relevant observations are collected in a pre-deployment setting, for instance through sandboxing, shadow mode, or controlled pilots, and organized into a labeled dataset \(\mathcal D_N\). The first decision gate asks whether empirical learning is \emph{warranted}. Relevant considerations include regime coverage, label stability, profile informativeness, and the feasibility of meaningful validation. If the answer is negative, the expert-defined rule remains operative, i.e., \(\mathsf{TW}^{\mathrm{op}}=\mathsf{TW}^{\mathrm{exp}}.\)
If learning is warranted, a candidate rule \(\widehat{\mathsf{TW}}\) is constructed using \(\mathcal D_N\) and evaluated. The second gate then asks whether this candidate is sufficiently adequate for governance use, considering factors such as validation performance, interpretability, regime coverage, and stability near boundaries. If the answer is positive, the operative rule becomes \(\mathsf{TW}^{\mathrm{op}}=\widehat{\mathsf{TW}};\) otherwise, \(\mathsf{TW}^{\mathrm{op}}=\mathsf{TW}^{\mathrm{exp}}.\) Therefore, whether learning or adoption is \emph{warranted} is a governance decision. For both gates, the applicable criteria and acceptance conditions should be specified ex ante, evaluated against  evidence, subject to  challenge, and evaluated by a governance authority. Table~\ref{tab:gate_warrant} summarizes the corresponding decisions, evidence, and controls.

%%%%%%%%%%%%%%%%%%%%%%%%%%%%%%%%%%%%%%%%%%%%%
% WARRANTING GATE 1 AND GATE 2
%%%%%%%%%%%%%%%%%%%%%%%%%%%%%%%%%%%%%%%%%%%%%
\begin{table*}[h!]
\centering
\small
\setlength{\tabcolsep}{4pt}
\renewcommand{\arraystretch}{1.10}

\begin{tabularx}{\linewidth}{
@{}
p{1.20cm}
p{3.0cm}
>{\raggedright\arraybackslash}X
>{\raggedright\arraybackslash}X
@{}
}
\toprule
\textbf{Gate}
&
\textbf{Decision}
&
\textbf{Warranting evidence}
&
\textbf{Governance controls}
\\
\midrule

\textbf{Gate 1}
&
\emph{Is empirical learning of \(\mathsf{TW}\) warranted?}
&
\(\cdot\) Governance value over the expert rule \newline
\(\cdot\) Sufficient regime coverage and profile variation \newline
\(\cdot\) Stable, documented labels \newline
\(\cdot\) Feasible independent validation
&
\(\cdot\) Criteria specified ex ante \newline
\(\cdot\) Evidence reviewed by domain and validation functions \newline
\(\cdot\) Decision and rationale approved and recorded by the governance authority
\\

\midrule

\textbf{Gate 2}
&
\emph{Is the learned candidate \(\widehat{\mathsf{TW}}\) warranted for adoption?}
&
\(\cdot\) Adequate validation and temporal generalization \newline
\(\cdot\) Coverage of rare or high-risk regimes \newline
\(\cdot\) Interpretability and boundary stability \newline
\(\cdot\) Structural admissibility and added governance value
&
\(\cdot\) Development and approval segregated \newline
\(\cdot\) Independent challenge against pre-specified criteria \newline
\(\cdot\) Candidate version, evidence, decision, and rationale recorded
\\

\bottomrule
\end{tabularx}

\caption{Warranting the two decision gates in Phase~2. Gate decisions are context-dependent governance judgments supported by documented evidence, ex ante criteria, independent challenge, and explicit decision authority.}
\label{tab:gate_warrant}
\end{table*}
%%%%%%%%%%%%%%%%%%%%%%%%%%%%%%%%%%%%%%%%%%%%%%%%%%%%%%%%%%%%%%%%%

\vspace{-2em}

\paragraph{Phase 3: Post-deployment monitoring and reassessment.}
After deployment, the operative rule \(\mathsf{TW}^{\mathrm{op}}\) is used to monitor trustworthiness over time. At review time \(t\), governance actors observe the trustworthiness profile \(\mathbf Q_A(t)\) and assign the corresponding level
\(\tau_A^{\mathrm{op}}(t):=\mathsf{TW}^{\mathrm{op}}\bigl(\mathbf Q_A(t)\bigr).
\)
These outputs support documentation, monitoring, and reassessment across the AI lifecycle. Depending on the operative rule, additional diagnostics can be reported to identify regime proximity and temporal change. (We will introduce two examples in Section~\ref{subsection:auditing_measures}.) When post-deployment evidence--e.g., target variable change through refinement of existing labels or changes in the criteria in which they are assigned--system updates, incidents, or other changes materially affect the monitored profile or the validity of the operative rule, reviewed cases may be sent to a further validation procedure and the governance process may return to Phase~1 or Phase~2. Post-deployment cases do not automatically update the dataset \(\mathcal D_N\). Any dataset extension or relabeling requires a documented validation procedure under the applicable protocol version. This is important because trustworthiness labels may evolve as the system, its environment, evidence base, or governance standard changes. Accordingly, both datasets and labeling procedures should be versioned. In regulated settings, such updates may also require review under the logic of \emph{substantial modification} in the AIA---see Art.~3(23), Art.~43(4), and Art.~72 \citep{EU_AI_Act_2024}. Finally, post-deployment monitoring is typically discrete and dimension-specific rather than continuous. The frequency with which \(\mathbf Q_A(t)\) can be updated depends on the evidence required to measure its components. Some dimensions may be refreshed frequently from logs or recent outputs, while others depend on delayed outcomes, periodic audits, or expert review. 
The governance procedure should therefore also document these measurement cadences and delays in evidence availability.

%%%%%%%%%%%%%%%%%%%%%%%%%%%%%%%%%%%%%%
%%%%%%%%%%%%%%%%%%%%%%%%%%%%%%%%%%%%%%
% AI GOVERNANCE WORKFLOW
%%%%%%%%%%%%%%%%%%%%%%%%%%%%%%%%%%%%%%%%%%%%%%%%%%%%%%

%%%%%%%%%%%%%%%%%%%%%%%%%%%%%%%%%%%%%%%%%
\subsection{Human Roles, Decision Rights, and Control Measures}
\label{subsection:human_roles_controls}
The implementation of the AI lifecycle governance procedure requires explicit allocation of human responsibilities and decision rights. Here, five functions should remain distinguishable: a system or product owner maintains the procedure and assembles the evidence; domain experts and representative users interpret the intended purpose, context of use, and consequences of trustworthiness failures; technical and data teams operationalize dimensions, maintain measurement pipelines, and implement candidate rules; an independent validation, risk, compliance, safety, or ethics function challenges the relevant assumptions and evidence; and a designated governance authority approves the operative rule \(\mathsf{TW}^{\mathrm{op}}\) and associated lifecycle actions. These functions should be supported by segregation of duties, independent review, documented escalation conditions, and human authorization for consequential governance decisions. The AI lifecycle governance procedure  therefore does not introduce a new governance body, as it can rely on organizational controls already used in model-risk management, quality assurance, safety governance, and regulated compliance. Its implementation may nevertheless remain resource-intensive because measurement, expert labeling, monitoring, and review require institutional capacity. Table~\ref{tab:human_governance_controls} summarizes a minimal allocation of responsibilities throughout the phases of the governance procedure and to promote assurance.

%%%%%%%%%%%%%%%%%%%%%%%%%%%%%%%%%%%%%%%%%
\subsection{Auditing Measures of the Operative Trustworthiness Level Function}
\label{subsection:auditing_measures}
Post-deployment auditing relies on the measured trustworthiness profile \(\mathbf Q_A(t)\in[0,1]^n\) and the operative trustworthiness level function \(\mathsf{TW}^{\mathrm{op}}\). Together they define the operative trustworthiness level trajectory
\(
\tau_A^{\mathrm{op}}(t):= \mathsf{TW}^{\mathrm{op}}\bigl(\mathbf Q_A(t)\bigr).
\)
The profile and level trajectories, together with marked lifecycle events such as \emph{shocks}, incidents, updates, or retraining, provide the basic reporting objects. Plotting these time series shows how the measured profile and the operative rule evolve over time. We complement these outputs with two diagnostics: boundary margin and profile drift.

\paragraph{Boundary margin.}
Let \(D=[0,1]^n\). Define
\[
\kappa_A^{\mathrm{op}}(t)
:=\inf_{\substack{\mathbf q\in D\\
\mathsf{TW}^{\mathrm{op}}(\mathbf q)\neq
\mathsf{TW}^{\mathrm{op}}(\mathbf Q_A(t))}}
d\!\left(\mathbf Q_A(t),\mathbf q\right).
\]
Thus, the boundary margin is the greatest lower bound of the distances from the current profile to all admissible profiles assigned a different trustworthiness level. A large value indicates that the current profile lies well inside its assigned regime, whereas a small value indicates proximity to a level transition. It is therefore a \emph{rule-relative measure of local fragility}. The distance function \(d\)
and any threshold used to interpret the margin are governance choices and should be documented.

\paragraph{Profile drift over time.}
For a documented horizon \(h>0\), define
\[
D_A(t;h)
:=
\frac{1}{\sqrt{n}}
\left\|
\mathbf Q_A(t)-\mathbf Q_A(t-h)
\right\|_2.
\]
Profile drift measures how far the system has moved in trustworthiness space over the preceding \(h\) time steps. Unlike the boundary margin, it does not depend on \(\mathsf{TW}^{\mathrm{op}}\) and does not by itself indicate whether the change is beneficial or harmful. 
The two diagnostics are therefore complementary: \(\kappa_A^{\mathrm{op}}(t)\) captures proximity to a governance boundary, whereas \(D_A(t;h)\) captures temporal movement in the underlying trustworthiness profile.

%%%%%%%%%%%%%%%%%%%%%%%%%%%%%%%%%%%%%%%%%%%%%
% HUMAN ROLES AND CONTROL MEASURES
%%%%%%%%%%%%%%%%%%%%%%%%%%%%%%%%%%%%%%%%%%%%%
\begin{table*}[h!]
\centering
\small
\setlength{\tabcolsep}{4pt}
\renewcommand{\arraystretch}{1.10}

\begin{tabularx}{\linewidth}{
@{}
p{1.4cm}
p{3.0cm}
>{\raggedright\arraybackslash}X
p{3.4cm}
@{}
}
\toprule
\textbf{Phase}
&
\textbf{Main actors}
&
\textbf{Responsibilities and controls}
&
\textbf{Required output}
\\
\midrule

\textbf{Phase 1}
&
\(\cdot\) System owner \newline
\(\cdot\) Domain experts \newline
\(\cdot\) Technical team \newline
\(\cdot\) Risk/compliance function
&
\(\cdot\) Define purpose, context, protocol, and level scale \newline
\(\cdot\) Specify labeling procedure and monitoring cadence \newline
\(\cdot\) Define escalation conditions \newline
\(\cdot\) Document expert agreement and independent challenge
&
\(\cdot\) Versioned protocol \newline
\(\cdot\) Labeling procedure \newline
\(\cdot\) Responsibility register
\\

\midrule

\textbf{Phase 2}
&
\(\cdot\) Data/model teams \newline
\(\cdot\) Domain experts \newline
\(\cdot\) Independent validation function
&
\(\cdot\) Assess label and data quality and regime coverage \newline
\(\cdot\) Conduct Gate~1 and Gate~2 assessments \newline
\(\cdot\) Evaluate validation, interpretability, and rule stability \newline
\(\cdot\) Preserve independent challenge and approval
&
\(\cdot\) Dataset record \newline
\(\cdot\) Validation and gate-decision record \newline
\(\cdot\) Operative-rule decision
\\

\midrule

\textbf{Phase 3}
&
\(\cdot\) Monitoring team \newline
\(\cdot\) System owner \newline
\(\cdot\) Operational users \newline
\(\cdot\) Review body
&
\(\cdot\) Monitor profiles, levels, and diagnostics \newline
\(\cdot\) Review incidents and system updates \newline
\(\cdot\) Authorize relabeling, restriction, rollback, or reassessment where required
&
\(\cdot\) Monitoring log \newline
\(\cdot\) Review decisions \newline
\(\cdot\) Updated governance objects
\\

\hline

\textbf{Assurance}
&
\(\cdot\) Internal audit or 
independent assurance function
&
\(\cdot\) Verify role assignment and segregation of duties \newline
\(\cdot\) Verify versioning and evidence retention \newline
\(\cdot\) Review escalation and corrective actions
&
\(\cdot\) Audit record \newline
\(\cdot\) Corrective-action plan
\\

\bottomrule
\end{tabularx}

\caption{Minimal allocation of human responsibilities and controls across the AI lifecycle governance procedure.}
\label{tab:human_governance_controls}
\end{table*}

\vspace{-2em}

%%%%%%%%%%%%%%%%%%%%%%%%%%%%%%%%%%%%%%%%
\subsection{A Simple Reporting Artifact}
\label{subsection:regulatory_tool}
Table~\ref{tab:checklist} in Appendix~\ref{appendix:minimal_artifact} provides a minimal artifact for documenting the proposed governance procedure. It records the trustworthiness protocol, level scale and labeling procedure, selection of the operative rule, current profile and diagnostics, governance responsibilities, version history, and escalation conditions. The artifact is initialized during Phases~1--2 and updated during post-deployment monitoring as new evidence, review decisions, or reassessment triggers arise. It thereby supports pre-deployment documentation, post-deployment monitoring, and reassessment while preserving an auditable record of how trustworthiness evidence is interpreted over time. These functions align with the documentation and lifecycle-governance logic of the AIA---see Art.~3(23), Art.~43, and Art.~72 \citep{EU_AI_Act_2024}.

%%%%%%%%%%%%%%%%%%%%%%%%%%%%%%%%%%%%%%%%
%%%%%%%%%%%%%%%%%%%%%%%%%%%%%%%%%%%%%%%%
\section{Illustrative AI Lifecycle Scenarios}
\label{section:empirical_study}
We illustrate the trustworthiness level empirical methodology through three synthetic lifecycle scenarios, which are not intended to establish external validity, showing how the proposed governance objects and decision gates behave under controlled variation in lifecycle dynamics, monitoring cadence, trustworthiness protocols, and expert-defined rules. The full  specifications of the scenarios and all figures reporting their results are presented in Appendix~\ref{appendix:sim_implementation}.

\vspace{-0.5em}

%%%%%%%%%%%%%%%%%%%%%%%%%%%%%%%%%%%%%%%%
\subsection{Scenario 1: Lifecycle Baseline}
\label{subsec:exp1_main}
The first, simple scenario considers two AI systems with the same two-dimensional trustworthiness protocol---accuracy and robustness---but different monitoring cadences. A \emph{recommender system} provides dense trustworthiness evidence, whereas a \emph{clinical decision-support system} is monitored more sparsely because some measures become available only after delay. Both trajectories include gradual degradation, adverse shocks, and corrective updates. The simulated shocks represent disruptions that may affect system trustworthiness, such as sudden distribution shifts, changes in user interaction patterns, or performance degradations revealed through delayed outcome review. The simulated updates represent stylized remediation actions, such as retraining, interface adjustments, or revised monitoring procedures. 
% scenario~1 provides the \emph{simplest} illustration of our methodology, showing how it can support trustworthiness monitoring.

Figures~\ref{fig:exp1_main_results} and~\ref{fig:exp1_boundary_drift} illustrate the resulting lifecycle patterns. In the dense recommender setting, the candidate trustworthiness level function reproduces the expert-defined levels closely and tracks transitions following shocks and updates. Boundary margins become small near regime transitions, while profile drift peaks around lifecycle events. The learned partition is simple. The sparse clinical trajectory exhibits coarser and noisier level dynamics; here the learned rule effectively compresses the expert-defined four-level scale into fewer empirically supported regimes. Here too, boundary margin is smallest near regime changes, while drift peaks at shocks and updates and remains noisier between events because measurements are sparse. The learned partition remains geometrically simple and captures the intended coarse ordering of the four clinical trustworthiness levels. Both learned candidate rules satisfy the coordinatewise monotonicity requirement imposed on trustworthiness level functions.
% The scenario therefore illustrates both the possibility of learning a compact operative rule and the need to evaluate whether loss of governance granularity is acceptable before adoption.

%%%%%%%%%%%%%%%%%%%%%%%%%%%%%%%%%%%%%%%
\subsection{Scenario 2: Single-System Lifecycle with Asynchronous Monitoring}
\label{subsec:exp2_main}
\begin{figure*}[h!]
\centering

\begin{minipage}[t]{0.95\textwidth}
    \centering
    \includegraphics[width=\linewidth]
    {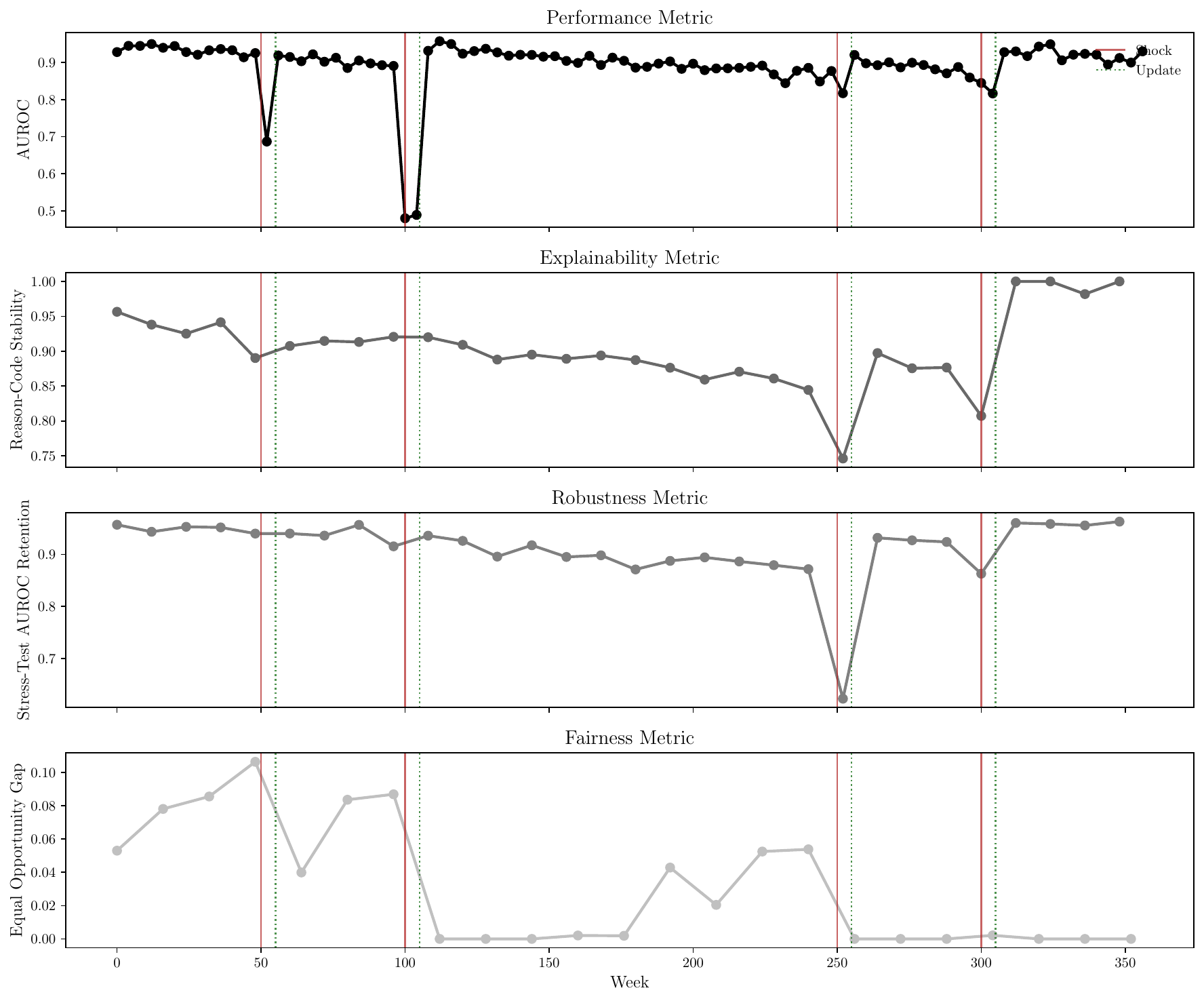}

    \smallskip
    \small
    (a) Asynchronously observed raw trustworthiness metrics
\end{minipage}
\par\vspace{0.8em}
\begin{minipage}[t]{0.95\textwidth}
    \centering
    \includegraphics[width=\linewidth]{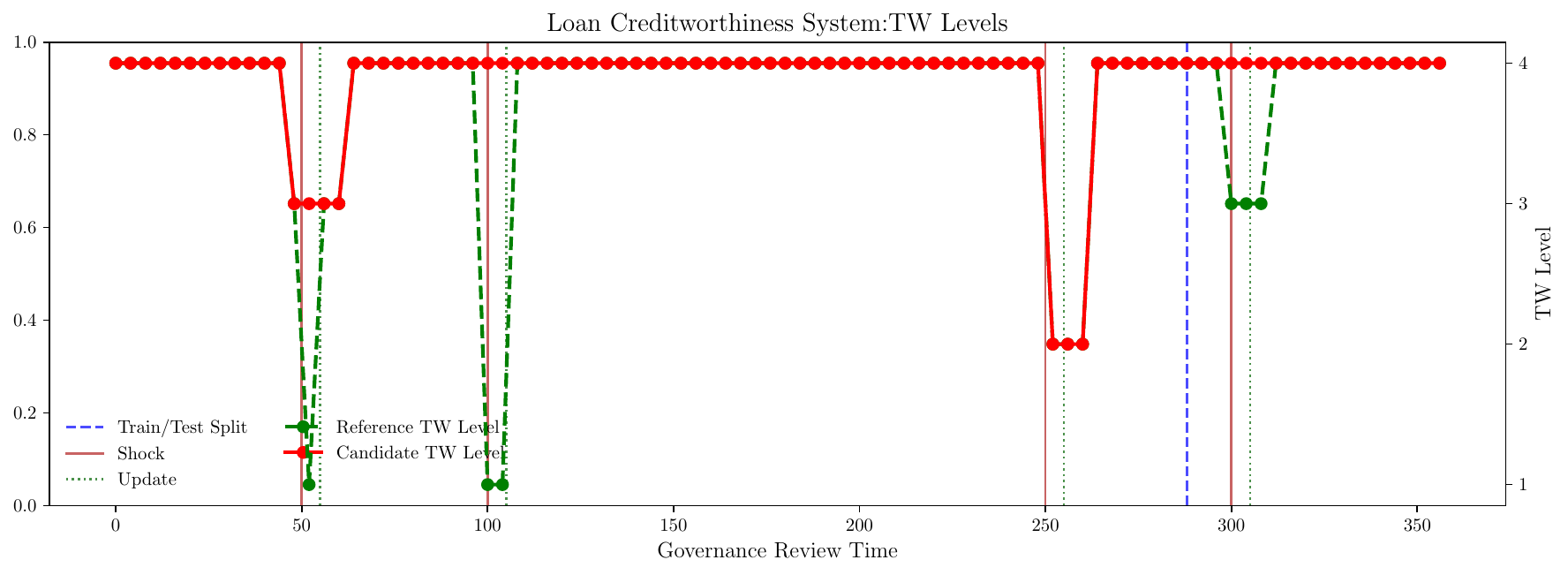}

    \smallskip
    \small
    (b) Latest-available trustworthiness profile and
    reference/predicted trustworthiness levels
\end{minipage}

\caption{Scenario 2: single-system lifecycle with asynchronous monitoring.
The upper panel reports the raw trustworthiness metrics observed at
different monitoring cadences. The lower panel reports the resulting
latest-available trustworthiness profiles and their reference and
predicted trustworthiness levels in the lending use case.
}
\label{fig:exp2_main_results}
\end{figure*}

Scenario~2 extends the baseline from Scenario~1 to a higher-dimensional profile while preserving the single-system setting. An AI system assesses the creditworthiness of loan applicants and the trustworthiness protocol consists of four dimensions: performance, explainability, robustness, and fairness. Each dimension is operationalized through a single metric with its own monitoring cadence. Then, the four time series are measured on different schedules and yield arrays of different lengths. Accordingly, trustworthiness levels are evaluated at discrete times, where each dimension is given by the latest available measurement. The dimension scores evolve under slow degradation, a small number of adverse shocks, and update events, i.e., corrective actions such as model revision or fairness interventions. Because of different measurement frequencies, their effects are not visible simultaneously across dimensions. Figures~\ref{fig:exp2_main_results} and ~\ref{fig:exp2_boundary_drift} show results. The learned rule largely tracks the dominant high-trustworthiness regime but compresses distinctions among weakly represented lower levels. During the training phase, despite slow degradation and adverse shocks, the system remains at the highest trustworthiness level for most of training, suggesting that  updates were effective at preserving the overall trustworthiness regime. The temporary drops from trustworthiness level~4 occur around the shock events simulating disruptions during development, pilot testing, or shadow-mode evaluation. Boundary margins also contract around several shock events and widen after subsequent updates. Held-out accuracy remains 0.83, while balanced accuracy falls to 0.50, reflecting substantial class imbalance and limited evidence for rarer regimes. The candidate also violates coordinatewise monotonicity, failing Gate~2. The expert-defined rule remains operative,
\(\mathsf{TW}^{\mathrm{op}}=\mathsf{TW}^{\mathrm{exp}}\).

%%%%%%%%%%%%%%%%%%%%%%%%%%%%%%%%%%%%%%%%%%%%%
% FIGURE: EXPERIMENT 2
%%%%%%%%%%%%%%%%%%%%%%%%%%%%%%%%%%%%%%%%%%%%%

%%%%%%%%%%%%%%%%%%%%%%%%%%%%%%%%%%%%%%%%
\subsection{Scenario 3: Governance-Context Comparison}
\label{subsec:exp3_main}
The third scenario compares two deployed copies of the same clinical decision-support system across two hospitals. The copies share intended purpose, trustworthiness protocol, and monitoring cadence, but differ in both the stability of their deployment environments and the granularity of the expert-defined trustworthiness level rules used by the local governance teams. The trustworthiness protocol consists of six dimensions: performance, calibration, robustness, fairness, explainability, and oversight quality. 
Hospital Alpha represents a stable environment governed by a coarse expert-defined rule. All observed profiles remain in trustworthiness level~4, so the AI lifecycle data contain no meaningful regime variation. Under Gate~1, empirical learning is therefore unwarranted and
\(
\mathsf{TW}^{\mathrm{op}}=\mathsf{TW}^{\mathrm{exp}}.
\)
Hospital Beta is exposed to stronger degradation, shocks, and corrective updates, and its expert labels span all four trustworthiness levels. Gate~1 therefore permits empirical learning, but the resulting candidate fails to preserve the refined governance scale: it does not recover the rare level~1 regime and its held-out accuracy falls to (0.50) while, differently from Scenario~2, it is coordinatewise monotone. This is a case of Gate~2 rejection; as a result one obtains
\(
\mathsf{TW}^{\mathrm{op}}=\mathsf{TW}^{\mathrm{exp}}.
\)

Figure~\ref{fig:exp3_main_results} and~\ref{fig:exp3_boundary_drift} illustrate these results. Hospital Alpha shows limited movement within the highest regime, whereas Hospital Beta exhibits substantially greater profile drift and the boundary  margin approaches zero at several points, indicating close proximity to candidate-level transitions. Further, for Hospital Beta the learned rule does not fully recover the refined expert scale: it performs reasonably on the training segment but deteriorates on the post-deployment segment, and it compresses the four-level reference scale into a coarser candidate rule. Thus, two hospitals deploying copies of the same AI system may differ in both governance procedures and the value of empirical learning. In the Alpha environment, learning adds little beyond the expert rule. In the Beta environment, learning is useful because its limitations reveal compression, boundary fragility, and post-deployment instability that warrant further review.

%%%%%%%%%%%%%%%%%%%%%%%%%%%
%%%%%%%%%%%%%%%%%%%%%%%%%%%
\section{Discussion and Conclusions}
\label{section:discussion_conclusions}
\emph{We introduced a lightweight methodology for auditable trustworthiness levels in AI lifecycle governance}. Its central object is the trustworthiness level function \(\mathsf{TW}\), which separates the multidimensional evidence represented by a trustworthiness profile of an AI system from the governance rule used to interpret that evidence over time. This function makes explicit a step that is often left implicit in trustworthiness assessment: multiple measured dimensions of AI functioning do not by themselves determine whether a system remains in an acceptable governance regime, approaches a boundary, or warrants review due, for instance, a substantial modification \citep{EU_AI_Act_2024}. Although \(\mathsf{TW}\) is a lossy summary, the underlying trustworthiness profile \(\mathbf{Q}_A(t)\) of the system \(A\) remains available for monitoring, diagnosis, audit over time.
At first sight, our methodology resembles established governance devices such as score-based algorithmic impact assessments, risk matrices, and risk-tiering or classification frameworks, all of which compress heterogeneous information into a smaller number of decision-relevant categories \citep{CanadaAIA2026,OECD2022AIClassification,cox2008risk,NISTAIRMF2023}.
Note that ordered levels also appear in recent FAccT work on responsible-AI maturity, where they characterize the extent to which organizations have developed and institutionalized responsible-AI practices \citep{reuel2025responsible}. 
Despite the similarity, the utility of our methodology lies in separating the underlying trustworthiness multi-dimensional evidence from the rule used to interpret it over time and treating that rule itself as a lifecycle governance object. The provenance of the rule can be expert-defined or empirically learned, while its adoption is governed through explicit decision gates. Furthermore, its level regions and transitions can be monitored over time and the operative rule can be versioned, challenged, retained, or replaced without discarding the underlying trustworthiness profile. As a result, \emph{we treated \(\mathsf{TW}\) as a key governed and interpretable governance object embedded in an AI lifecycle procedure}. The distinction among \(\mathsf{TW}^{\mathrm{exp}}\), a candidate \(\widehat{\mathsf{TW}}\), and the operative rule \(\mathsf{TW}^{\mathrm{op}}\) makes empirical learning conditional upon two distinct governance gates.  The illustrative scenarios showed why these gates should remain separate: learning may recover a useful compact rule, add little because observed regimes lack variation, or produce a candidate whose compression of rare states warrants rejection. In summary, our methodology offers two main contributions. First, it makes trustworthiness judgments more explicit and contestable: multiple dimensions, measures, aggregation choices, labeling procedures, learned rules, and monitoring outputs can all be documented and revised over time. Second, it connects pre-deployment governance with post-deployment monitoring by converting a comparatively static design-time governance standard into a dynamic monitoring layer using \(\mathsf{TW}\) that can be repeatedly applied to new trustworthiness profiles. The operative  rule \(\mathsf{TW}^{\mathrm{op}}\) provides a contestable evidence-based layer that can inform conformity assessment, post-market monitoring, and reassessment by making level assignments, transitions, margins, drift, and protocol updates explicit over time for different types of AI systems \citep{EU_AI_Act_2024}.

A few possible criticisms deserve clarification. First, \emph{the formalization is algorithmically lightweight.} Our use of decision trees is a proof-of-concept choice, as  interpretable threshold-based rules can facilitate documentation and review, but predictive performance alone is insufficient for adoption. Candidate rules must also preserve relevant regime distinctions and satisfy context-appropriate requirements concerning interpretability, temporal generalization, boundary stability, and structural consistency. More expressive model classes may be appropriate where trustworthiness regions are high-dimensional or interaction-heavy, provided that their use remains compatible with governance requirements.

Second, 
\emph{the formalization does not eliminate the need for expert judgment in assigning trustworthiness labels.} This is not peculiar to our proposal, but a standard feature of governance-oriented assessment in domains where acceptable, borderline, and unacceptable system states must be determined under context-specific normative and operational constraints. Our methodology assumes that reference labels are produced under a documented adjudication procedure that is specific to the relevant institution, domain, and governance setting. In other words, it does not attempt to standardize such procedures across all contexts. These procedures should be embedded in existing audit, assurance, risk-management, or compliance procedures appropriate to the AI system, as making trustworthiness judgments explicit and contestable is not sufficient for legitimacy at the institutional level.

Third, \emph{the methodology is only as informative as the trustworthiness object that an institution can actually construct and maintain.} In real-world settings, the main difficulty may lie less in learning the trustworthiness level rule than in defining a meaningful protocol, aggregating heterogeneous measures, obtaining sufficiently informative profiles over time, and producing stable reference labels. Where dimensions are weakly measured, labels are sparse or inconsistent, classes are highly imbalanced, or protocol versions drift over time, the learned rule may become coarse, unstable, or effectively uninformative. The methodology is therefore best understood as an institutionally dependent governance layer: it becomes actionable only when the underlying measurement, aggregation, labeling, and review procedures are themselves documented and sufficiently mature.  Future work should test the methodology on real domain-specific cases, compare decision trees with other interpretable model classes, and refine validation and relabeling procedures under institution-specific protocols. Further work should also study robustness under label noise and protocol drift, and extend the approach to multi-agent settings and institutional workflows.

Finally, \emph{our illustrative scenarios are synthetic} and show the  decision logic of the methodology rather than its effectiveness on real-world data. Future work should evaluate the procedure in domain-specific deployments, study how organizations establish and contest gate criteria, and examine robustness under measurement uncertainty, rare regimes, label disagreement, and evolving protocols. The methodology discussed in this work has shown how governance actors can tackle these challenges using the trustworthiness level function as a governance instrument  supporting documentation, monitoring, escalation, and reassessment.

%%%%%%%%%%%%%%%%%%%%%%%%%
\section*{Acknowledgments}
We acknowledge partial support by the Swiss National Science Foundation (SNSF), grant no. 229061.

%%%%%%%%%%
\clearpage
\appendix

%%%%%%%%%%%%%%%%%%%%%%%%%%%%%%%%%%%%%%%%%%%%
\section{Trustworthiness dimensions, their measures and collection over time.}
\label{appendix:table_TW}
%%%%%%%%%%%%%%%%%
% ORIGINAL TABLE
%%%%%%%%%%%%%%%%%
\begin{table*}[h!]
\scriptsize
\centering
\caption{Representative trustworthiness dimensions, example measures, and their post-deployment collection logic, adapted from  \citet{kemmerzell2025towards}.}
\begin{tabularx}{\linewidth}{@{}>{\raggedright\arraybackslash}p{3.5cm}p{5.5cm}X@{}}
\toprule
\textbf{Dimension} & \textbf{Example measures} & \textbf{Typical post-deployment collection / update logic} \\
\midrule

\textbf{Accuracy/predictive performance}
& Squared error; F1 score; AUROC. 
& Usually updated on recent resolved cases only. It is therefore often delayed and estimated over rolling validation windows. \\

\textbf{Calibration / uncertainty}
& Brier score; expected calibration error.
& Usually updated only when outcomes become available, so it is typically assessed on delayed validation windows rather than live logs. \\

\textbf{Fairness}
& Demographic parity; disparate impact; equalized odds; false positive error-rate balance; counterfactual fairness
& Often evaluated on batches of recent labeled cases, since many metrics require protected attributes and realized outcomes. \\

\textbf{Transparency / explainability}
& Faithfulness; robustness of explanations; explanation complexity; task performance; documentation review.
& Some indicators can be updated directly from model behavior or documentation, while human-centered ones are usually assessed periodically. \\

\textbf{Privacy and data governance}
& \(k\)-anonymity; \(t\)-closeness; differential privacy \((\epsilon,\delta)\); membership-inference attack success rate.
& Commonly assessed through periodic audits, attack simulations, or after data/model updates rather than continuously. \\

\textbf{Technical robustness and safety}
& Performance drop under shift; attack success rate.
& Drift indicators may be monitored semi-continuously, whereas robustness and safety are often checked through periodic tests or red-teaming. \\

\textbf{Human agency and oversight}
& Override rate; escalation rate; task completion time; error rate; user satisfaction; reliance calibration.
& Operational indicators can often be logged continuously, while user-centered measures are usually updated through periodic review or feedback collection. \\

\textbf{Societal/environmental well-being}
& CO\(_2\) footprint of training and operation; number of model parameters; model runtime.
& Typically documented through reporting cycles, retraining events, or sustainability audits. \\

\bottomrule
\end{tabularx}
\label{table:Q_dimensions}
\end{table*}

%%%%%%%%%%%%%%%%%%%%%%%%%%%%%%%%%%%%%%%%
%%%%%%%%%%%%%%%%%%%%%%%%%%%%%%%%%%%%%%%%
\section{A Simple Reporting Artifact}
\label{appendix:minimal_artifact}
Table~\ref{tab:checklist} shows a simple governance artifact for documenting how trustworthiness levels are defined, learned, and monitored over time that can accompany the proposed methodology of AI governance. The artifact is embedded in the AI lifecycle governance procedure as follows. First, it is completed in pre-deployment settings (Phases 1--2). Then, it is updated post-deployment (Phase 3), when new measurements, delayed outcomes, review decisions, and reassessment triggers arise. 
By recording design choices, the learned rule, and lifecycle monitoring information, the artifact supports three functions. It supports \emph{pre-deployment documentation} by making explicit how trustworthiness is defined, measured, and labeled before the empirical rule is learned. This is aligned with the conformity assessment documentation in the AIA---see Art.~43. It supports \emph{post-deployment monitoring} by providing a compact schema for reporting the system's current trustworthiness profile, level, and diagnostics over time. Finally, it supports \emph{reassessment} by making review conditions, rule updates, and possible changes in the labeling protocol more transparent and contestable. These last two functions align with the governance of post-deployment monitoring as per the AIA---see Art.~3(23), Art.~43(4), and Art.~72.

%%%%%%%%%%%%%%%%%%%%%%%%%%%%%%%%
\begin{table}[t]
\centering
\small
\setlength{\tabcolsep}{4pt}
\renewcommand{\arraystretch}{1.08}
\begin{tabularx}{\columnwidth}{@{}p{3.5cm}X@{}}
\toprule
\textbf{Item} & \textbf{Minimal content} \\
\midrule
\textbf{System and context} &
Intended purpose, context of use, operational constraints, excluded uses. \\

\textbf{Trustworthiness protocol} &
Selected dimensions, inclusion rationale, omitted dimensions where relevant, hard/soft dimensions if used. \\

\textbf{Measures and aggregation} &
Operational definitions, normalization, aggregation choices, data sources, measurement cadence, evidence windows, limitations. \\

\textbf{Level scale and labels} &
Scale $1,\ldots,K$, level interpretation, labeling procedure, audit-label criteria, protocol version. \\

\textbf{Learned rule and validation} &
Model family, hyperparameter selection, complexity control, validation strategy, learned tree or equivalent rule. \\

\textbf{Learning value over expert rule} &
Whether learning improves, compresses, confirms, or fails to justify replacing the expert-defined rule. \\

\textbf{Data and label adequacy} &
Regime coverage, class balance, missingness, delayed outcomes, measurement uncertainty, label disagreement, protocol drift. \\

\textbf{Current monitoring state} &
Current profile, predicted level, boundary margin, profile drift, marked shocks, updates, or retrainings. \\

\textbf{Governance responsibility} &
Actor or unit responsible for maintaining the artifact, approving protocol changes, and escalating review. \\

\textbf{Update and evidence record} &
Update cadence; logs, validation reports, audit decisions, expert-review notes, and protocol/version history. \\

\textbf{Escalation and reassessment} &
Conditions for no action, intensified monitoring, expert review, relabeling, retraining, documentation update, or reassessment. \\
\bottomrule
\end{tabularx}
\caption{Minimal governance artifact for documenting trustworthiness level definition, learning, monitoring, and reassessment across the three AI lifecycle governance phases.}
\label{tab:checklist}
\end{table}

%%%%%%%%%%%%%%%%%%%%%%%%%%%%%%%%%%%%%%%%%%%
\section{Simulation Study: Details on the Implementation}
\label{appendix:sim_implementation}

\paragraph{Common Simulation Logic for Trustworthiness Dimensions.} Across the experiments, each trustworthiness-relevant quantity is generated from the same basic lifecycle model. Writing \(z(t)\) for the raw or latent value of one dimension at time \(t\), we simulate
\[
z(t)
=
\operatorname{clip}_{[0,1]}\left( b-d\,\tau_t+\varepsilon^{(1)}(t)
+\sum_j s_j\,\mathbf 1\{t\ge t_j^{\mathrm{shock}}\}
+\sum_k u_k\,\mathbf 1\{t\ge t_k^{\mathrm{update}}\}
+\varepsilon^{(2)}(t)
\right),
\]
where \(b\) is a baseline level, \(d\) a drift coefficient, \(\tau_t=t/(T-1)\) a normalized time index, \(s_j\) and \(u_k\) are the magnitudes of shocks and updates, and \(\operatorname{clip}_{[0,1]}\) truncates the result to the unit interval. The term \(b-d\tau_t+\varepsilon^{(1)}(t)\) represents a noisy drift component, namely a slowly degrading baseline trajectory perturbed by small process-level fluctuations. The second noise term \(\varepsilon^{(2)}(t)\) represents additional observation-level variability before clipping. Both noise terms follow a normal distribution with zero mean; their standard deviations differ across experiments.

The generated quantity \(z(t)\) is first interpreted as a raw or normalized
metric according to the convention of each scenario. In Scenario~1, accuracy
and robustness are already oriented so that higher values indicate greater
trustworthiness and therefore enter the trustworthiness profile directly.
In Scenarios~2 and~3, performance, explainability, robustness, and the other
quality measures are likewise normalized in the higher-is-better direction.
Fairness is initially simulated as an Equal Opportunity Gap
\(g_{\mathrm{fair}}(t)\in[0,1]\), for which lower values indicate better
fairness. Before entering the trustworthiness profile, it is therefore
re-oriented as
\[
Q_{\mathrm{fair}}(t)
:=
1-g_{\mathrm{fair}}(t).
\]
Accordingly, all coordinates \(Q_i(t)\) entering the trustworthiness profiles
in the three scenarios satisfy the higher-is-better convention adopted in
Section~\ref{subsection:protocol_profile_levels}.

\paragraph{Explicit reference rules.} The explicit reference rules reported below are used to generate reproducible synthetic expert labels. They represent stylized outputs of an expert adjudication process; their closed-form specification is provided for simulation reproducibility and should not be interpreted as implying that an equally inexpensive operational rule is available in practice.

\paragraph{Common decision-tree learning procedure.} Candidate trustworthiness level learning is implemented as an ordinal
multi-class classification task. The labels \(Y_i\) are discrete trustworthiness levels, ordered from lower to higher trustworthiness. The recommender case in scenario~1 uses three levels, whereas the clinical case in scenario~1, scenario~2, and Hospital Beta in scenario~3 use four levels. Hospital Alpha in scenario~3 is the exception: because all observations remain in level~4 under the local expert rule, no empirical classifier is learned and the expert-defined rule remains operative.

For each learned rule, the available data are split chronologically into a training segment containing the first \(80\%\) of observations and a test segment containing the remaining \(20\%\), so that evaluation preserves temporal order. On the training segment, we fit a \texttt{DecisionTreeClassifier} and tune its main structural hyperparameters by temporal cross-validation using \texttt{TimeSeriesSplit}. The tree itself is trained as a classification model, using either the \texttt{gini} or \texttt{entropy} split criterion. Because the labels are ordered, model selection is driven primarily by mean absolute error over the predicted and reference levels, which treats errors between adjacent levels as less severe than errors across more distant levels. Accuracy, balanced accuracy, and macro-F1 are also recorded as complementary multi-class performance measures. The common search grid varies the split criterion, the maximum number of leaves, the maximum tree depth, and the cost-complexity pruning parameter \(\texttt{ccp\_alpha}\). Table~\ref{tab:dt_grid} reports the common hyperparameter grid used across the learned experiments.

\begin{table}[h!]
% \scriptsize
\centering
\caption{Common decision-tree hyperparameter grid used for all learned trustworthiness level rules.}
\label{tab:dt_grid}
\begin{tabular}{ll}
\toprule
\textbf{Hyperparameter} & \textbf{Candidate values} \\
\midrule
\texttt{criterion} & \(\{\texttt{gini},\texttt{entropy}\}\) \\

\texttt{max\_leaf\_nodes} & \(\{3,4,5,6,8,10,12,\texttt{None}\}\) \\

\texttt{max\_depth} & \(\{2,3,4,5,6,8,\texttt{None}\}\) \\

\(\texttt{ccp\_alpha}\) & \(\{0,10^{-4},10^{-3},10^{-2},10^{-1}\}\) \\
\bottomrule
\end{tabular}
\end{table}

After hyperparameter tuning, the best tree is refit on the full training segment and treated as a candidate trustworthiness level function
\(\widehat{\mathsf{TW}}\) for subsequent governance evaluation.

\paragraph{Boundary margin for the candidate rules.} 
For candidate rules evaluated at Gate~2, we analogously define \(\widehat{\kappa}_A(t)\) by replacing
\(\mathsf{TW}^{\mathrm{op}}\) with \(\widehat{\mathsf{TW}}\) in the boundary-margin definition above. In all simulations,
\(d\) is the Euclidean distance on \([0,1]^n\).

%%%%%%%%%%%%%%%%%%%%%%%%%%%%%%%%%%%%%%%%%%%
\subsection{Scenario 1}
\label{app:exp_1}
\begin{figure}[h!]
\centering

%========================
% Panel (a): Boundary distance
%========================
\includegraphics[width=0.8\columnwidth]{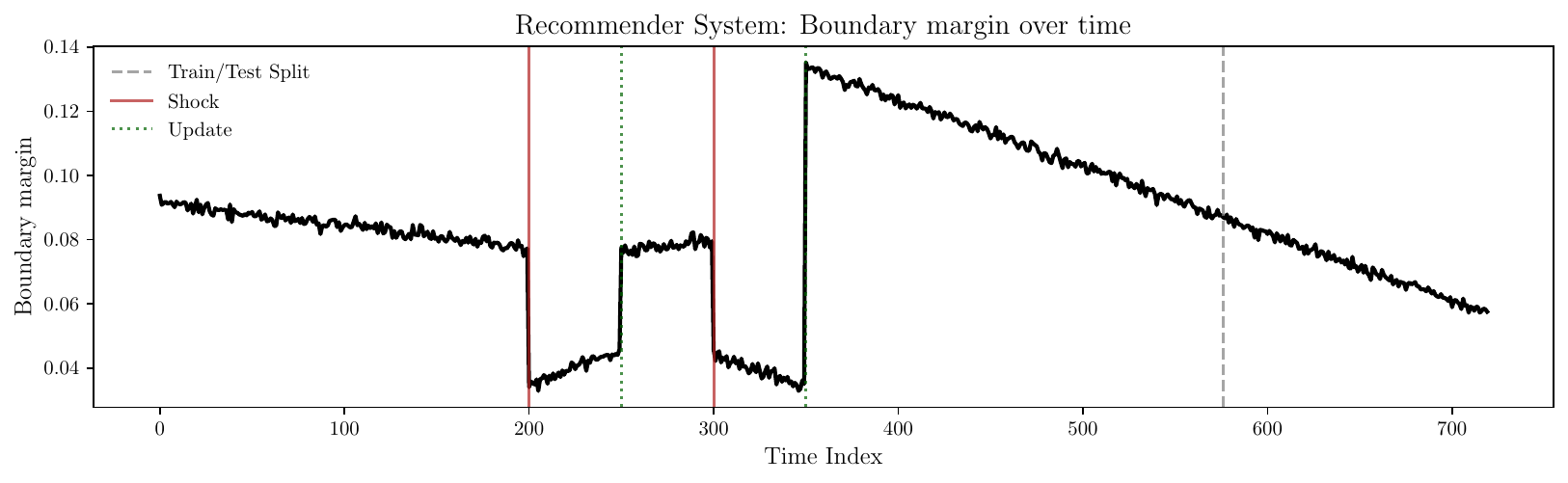}

\small (a) Recommender system: Boundary margin over time 

\includegraphics[width=\columnwidth]{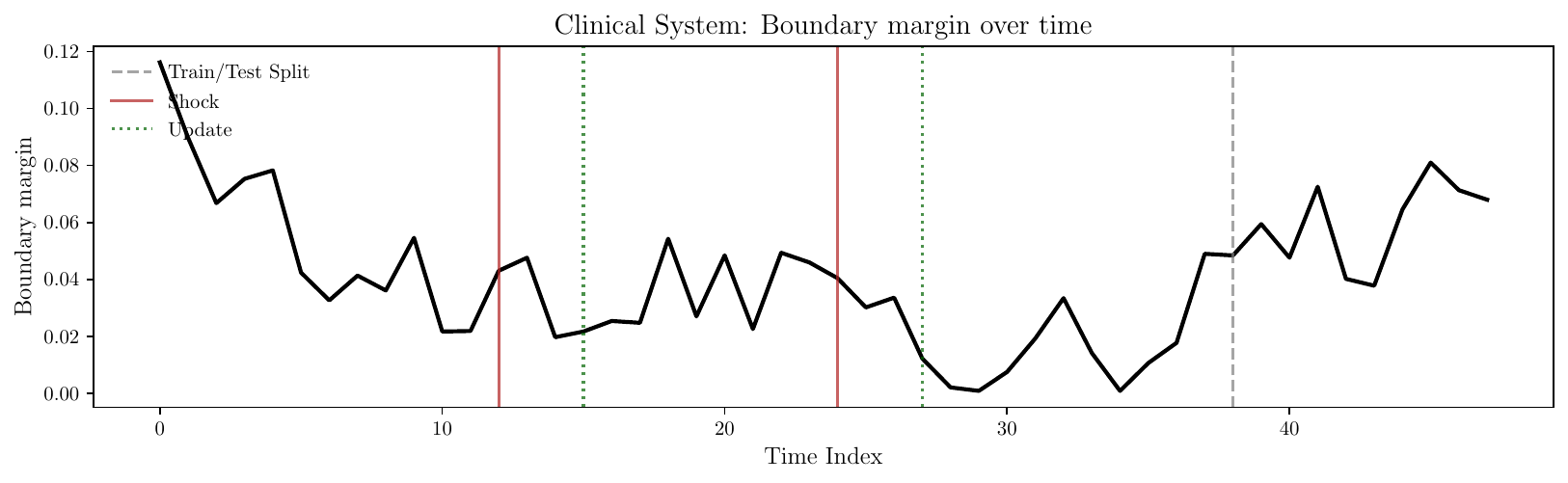}

\small (b) Clinical decision-support system: Boundary margin over time

\vspace{0.6em}

%========================
% Panel (b): Profile drift
%========================
\includegraphics[width=\columnwidth]{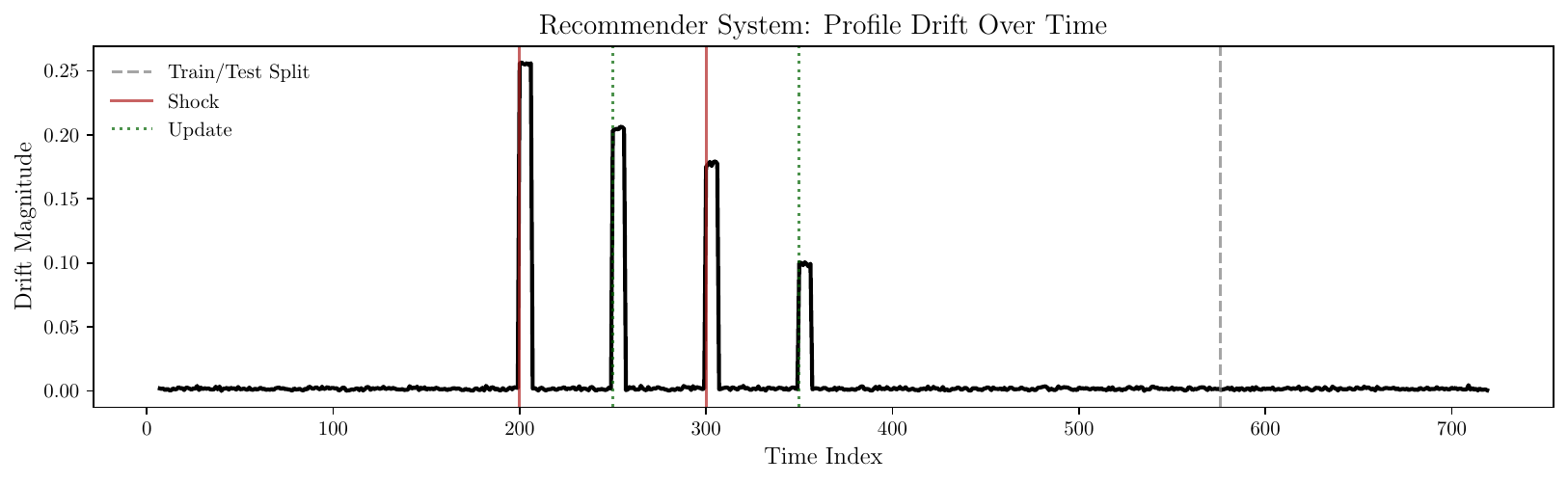}

\small (c) Recommender system: Profile drift over time

\includegraphics[width=\columnwidth]{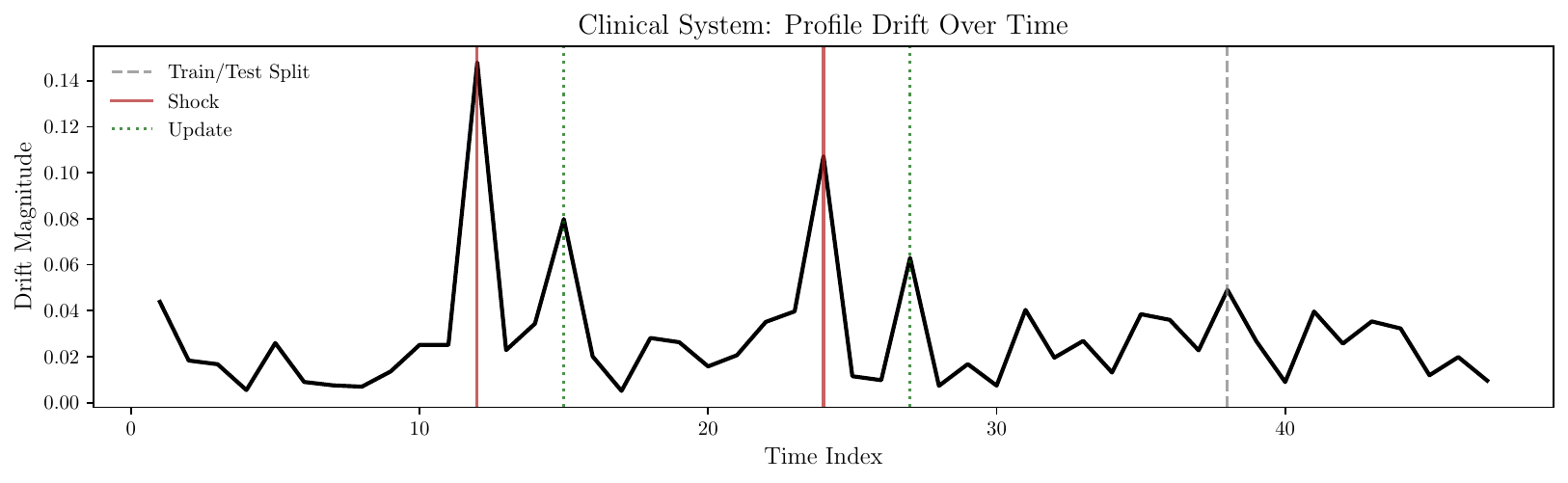}

\small (d) Clinical decision-support system: Profile drift over time

\caption{Scenario 1. Reporting boundary margin and profile drift over time.}
\label{fig:exp1_boundary_drift}
\end{figure}

This appendix provides the technical details of scenario~1, the two-dimensional lifecycle baseline introduced in Section~\ref{subsec:exp1_main}, and displays boundary margin and profile drift over time in Figure~\ref{fig:exp1_boundary_drift}. For the profile-drift diagnostic, we use a seven-step horizon for the dense recommender trajectory and a one-step horizon for the sparse clinical trajectory. Scenario~1 uses two directly measured trustworthiness dimensions, accuracy and robustness, with no within-dimension aggregation.

\paragraph{Simulation design.} The recommender trajectory is generated at a dense monitoring cadence, whereas the clinical trajectory is generated at a sparse cadence; both follow the common simulation logic above, but differ in step count, event timing, and expert-defined trustworthiness scale. Table~\ref{tab:exp1_design} summarizes the main design choices. The two noise terms use the \(\texttt{noise\_std}\) parameter specified for each trajectory. For the dense recommender trajectory, we set \(\texttt{noise\_std}=0.001\), so the process-level perturbation has standard deviation \(0.0005\) and the observation-level perturbation has standard deviation \(0.001\). For the sparse clinical trajectory, we set \(\texttt{noise\_std}=0.02\), so the corresponding standard deviations are \(0.01\) and \(0.02\), respectively.

\begin{table*}[h!]
\scriptsize
\centering
\caption{Scenario~1: simulation design summary. The recommender system is simulated over \(720\) dense daily-like observations; shocks take place at steps \((200,300)\) and updates at  \((250,350)\).  The clinical system is simulated over \(48\) sparse monthly-like observations; shocks take place at steps \((12,24)\) and updates at  \((15,27)\).}
\label{tab:exp1_design}
\begin{tabular}{l l r r l r r r}
\toprule
\textbf{System} & \textbf{Dimension} & \textbf{Baseline \(b\)} & \textbf{Drift \(d\)} & \textbf{Cadence} & \textbf{Obs.}  & \textbf{Shock effects} &  \textbf{Update effects}  \\
\midrule
Recommender & Accuracy   & 0.80 & 0.15 & dense / daily-like   & 720 &  \(-(0.20,0.15)\) &  \((0.25,0.10)\)  \\
Recommender & Robustness & 0.65 & 0.05 & dense / daily-like   & 720 &  \(-(0.30,0.20)\) & \((0.15,0.10)\)  \\
Clinical    & Accuracy   & 0.90 & 0.18 & sparse / monthly-like & 48  &  \(-(0.15,0.10)\) &  \((0.10,0.10)\)  \\
Clinical    & Robustness & 0.85 & 0.16 & sparse / monthly-like & 48    & \(-(0.10,0.10)\) & \((0.05,0.05)\)  \\
\bottomrule
\end{tabular}
\end{table*}

\paragraph{Expert-based trustworthiness level rules.} Table~\ref{tab:exp1_expert_rules} reports the expert-defined trustworthiness level rules used to generate the reference labels in scenario~1. The recommender and clinical settings use different governance scales, reflecting their different intended purposes and risk profiles.

\begin{table}[h!]
\centering
\caption{Scenario~1: reference trustworthiness level rules used to generate
the synthetic expert labels.}
\label{tab:exp1_expert_rules}
\begin{tabular}{llll}
\toprule
\textbf{System} & \textbf{Level} & \textbf{Rule} & \textbf{Interpretation} \\
\midrule
Recommender
& 1
& \(Q_{\mathrm{acc}} < 0.60\)
& risky \\

Recommender
& 2
& \(Q_{\mathrm{acc}} \ge 0.60\) and \(Q_{\mathrm{rob}} < 0.60\)
& acceptable \\

Recommender
& 3
& \(Q_{\mathrm{acc}} \ge 0.60\) and \(Q_{\mathrm{rob}} \ge 0.60\)
& good \\
\midrule

Clinical
& 1
& \(\min(Q_{\mathrm{acc}},Q_{\mathrm{rob}}) < 0.65\)
& risky \\

Clinical
& 2
& \(0.65 \le \min(Q_{\mathrm{acc}},Q_{\mathrm{rob}}) < 0.75\)
& acceptable \\

Clinical
& 3
& \(0.75 \le \min(Q_{\mathrm{acc}},Q_{\mathrm{rob}}) < 0.85\)
& good \\

Clinical
& 4
& \(\min(Q_{\mathrm{acc}},Q_{\mathrm{rob}}) \ge 0.85\)
& excellent \\

\bottomrule
\end{tabular}
\end{table}

\paragraph{Learning results.} The learned candidate rules are also structurally admissible with respect to coordinatewise monotonicity. Up to the displayed numerical precision of the learned thresholds, the recommender tree assigns level~1 when \(Q_{\mathrm{rob}}\leq 0.56\) and \(Q_{\mathrm{acc}}\leq 0.59\), level~2 when \(Q_{\mathrm{rob}}\leq 0.56\) and \(Q_{\mathrm{acc}}>0.59\), and level~3 when \(Q_{\mathrm{rob}}>0.56\). Hence, increasing \(Q_{\mathrm{acc}}\) can only move the assignment from level~1 to level~2, while increasing \(Q_{\mathrm{rob}}\) across its threshold can only move it from level~1 or~2 to level~3. The clinical tree depends only on robustness: it assigns level~1 for \(Q_{\mathrm{rob}}\leq 0.65\), level~2 for \(0.65<Q_{\mathrm{rob}}\leq 0.77\), and level~3 for \(Q_{\mathrm{rob}}>0.77\), while remaining invariant to changes in \(Q_{\mathrm{acc}}\). Consequently, for both candidate rules, \(\mathbf q\preceq\mathbf q'\) implies \(\widehat{\mathsf{TW}}(\mathbf q)\leq\widehat{\mathsf{TW}}(\mathbf q')\).

Table~\ref{tab:exp1_model_results} shows that, in the dense recommender setting, the learned trustworthiness rule reproduces the expert-based levels perfectly on both the training and test segments. In the sparse clinical setting, performance remains strong overall, with near-perfect training accuracy and perfect test accuracy, while balanced accuracy and macro-F1 on the training segment are lower because some trustworthiness levels are only weakly represented. In both cases, temporal cross-validation selected the same shallow tree, consistent with the role of Scenario~1 as an intentionally simple and
interpretable baseline. In the sparse clinical setting, the reference governance scale contains four levels, whereas the candidate tree recovers
only three empirically supported regimes. This illustrates that, under strong
simplicity constraints and limited support for one regime, a candidate rule may compress rather than preserve the full granularity of the reference scale. Whether such compression is acceptable is therefore a Gate~2 governance question rather than a consequence of predictive performance alone.

\begin{table*}[h!]
% \scriptsize
\centering
\caption{Scenario~1: predictive performance of the learned trustworthiness level rules. In both settings, temporal cross-validation selected the same simple tree configuration (\texttt{gini}, \texttt{max\_depth}=2, \texttt{max\_leaf\_nodes}=3, \(\texttt{ccp\_alpha}=0\)). Abbreviations: tr=train, te=test, Acc=accuracy, Bal.Acc=balanced accuracy.}
\label{tab:exp1_model_results}
\begin{tabular}{lrrrr}
\toprule
\textbf{System} & \textbf{MAE (tr / te)} & \textbf{Acc. (tr / te)} & \textbf{Bal. Acc. (tr / te)} & \textbf{Macro-F1 (tr / te)} \\
\midrule
Recommender & 0.00 / 0.00 & 1.00 / 1.00 & 1.00 / 1.00 & 1.00 / 1.00 \\
Clinical   & 0.05 / 0.00 & 0.95 / 1.00 & 0.75 / 1.00 & 0.73 / 1.00 \\
\bottomrule
\end{tabular}
\end{table*}

%%%%%%%%%%%%%%%%%%%%%%%%%%%%%%%%%%%%%%%%%%%%%
% FIGURE: EXPERIMENT 1
%%%%%%%%%%%%%%%%%%%%%%%%%%%%%%%%%%%%%%%%%%%%%
\begin{figure*}[h!]
\centering

\begin{minipage}[t]{0.85\textwidth}
    \centering
    \includegraphics[width=\linewidth]
    {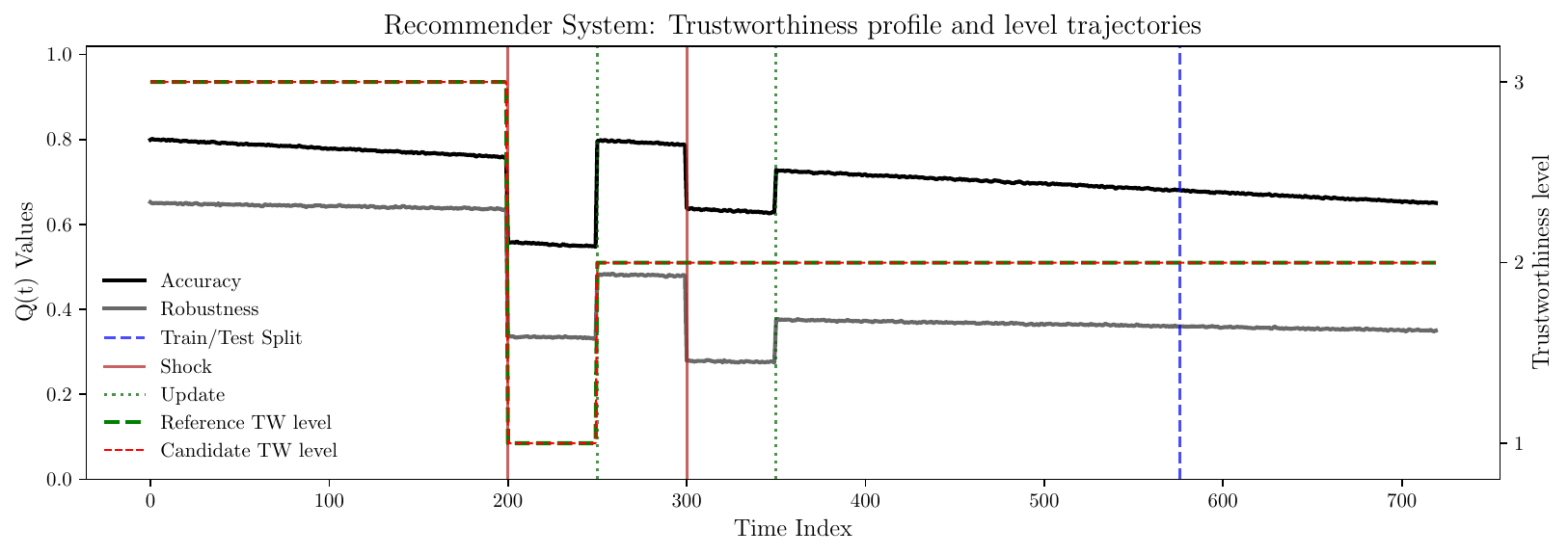}

    \smallskip
    \small
    (a) Recommender system: trustworthiness profile and
    reference/predicted trustworthiness levels
\end{minipage}
\par\vspace{0.8em}
\begin{minipage}[t]{0.85\textwidth}
    \centering
    \includegraphics[width=\linewidth]{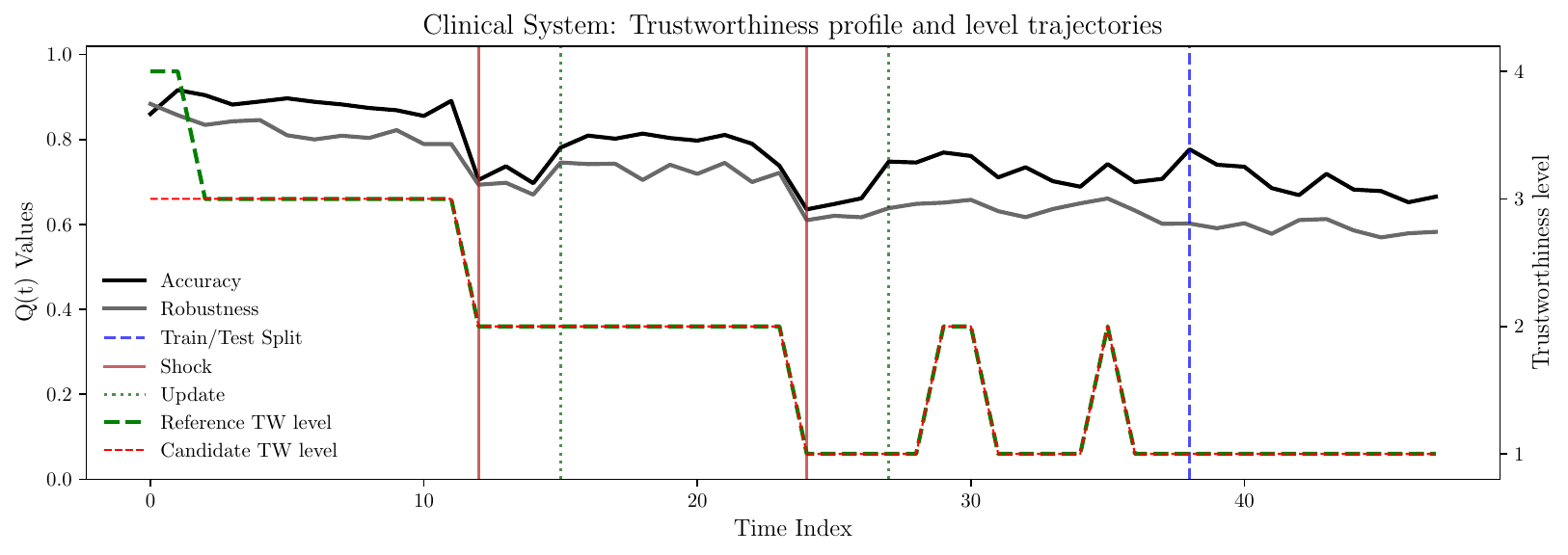}

    \smallskip
    \small
    (b) Clinical decision-support system: trustworthiness profile and
    reference/predicted trustworthiness levels
\end{minipage}

\caption{Scenario 1: lifecycle baselines.
Trustworthiness-profile and level trajectories for a densely monitored recommender system and a sparsely monitored clinical decision-support system. The two systems use the same two-dimensional profile structure but differ in monitoring cadence, lifecycle dynamics, and expert-defined trustworthiness scales.
}
\label{fig:exp1_main_results}
\end{figure*}

%%%%%%%%%%%%%%%%%%%%%%%%%%%%%%%%%%%%%%%%%%%%%
% SINGLE-COLUMN FIGURE: BOUNDARY DISTANCE + PROFILE DRIFT
\newpage

%%%%%%%%%%%%%%%%%%%%%%%%%%%%%%%%%%%%%%%%%%%
\subsection{Scenario 2}
\label{app:exp_2}
\begin{figure}[h!]
\centering

%========================
% Panel (a): Boundary distance
%========================
\includegraphics[width=\columnwidth]{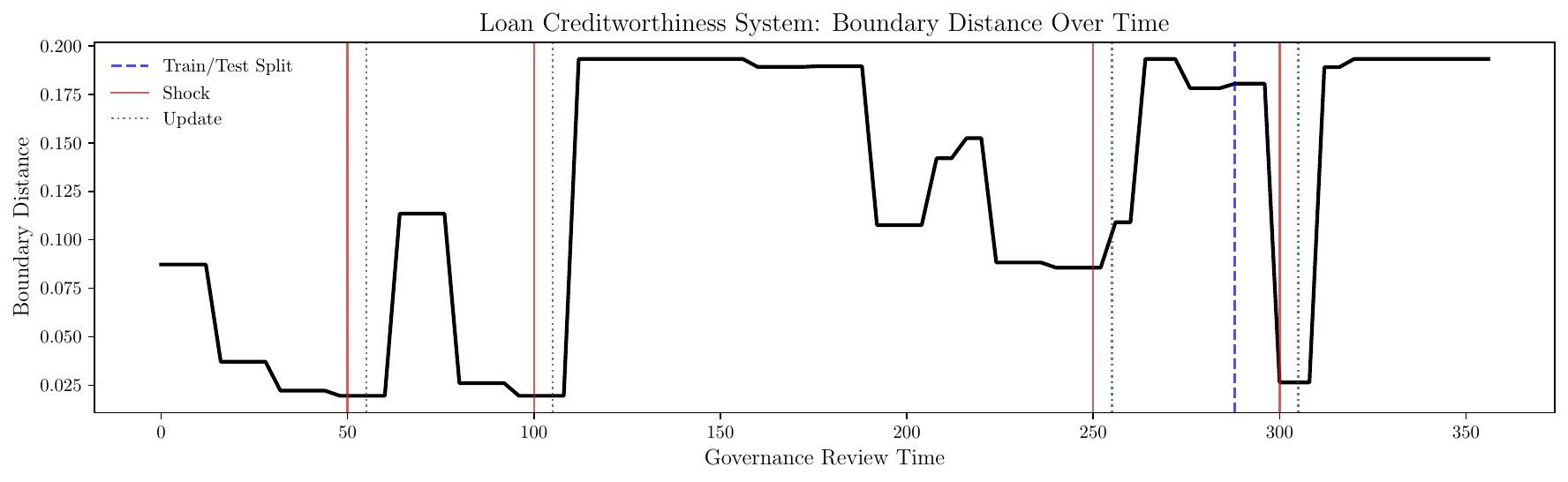}

\small (a) Boundary margin over time

\vspace{0.6em}

%========================
% Panel (b): Profile drift
%========================
\includegraphics[width=\columnwidth]{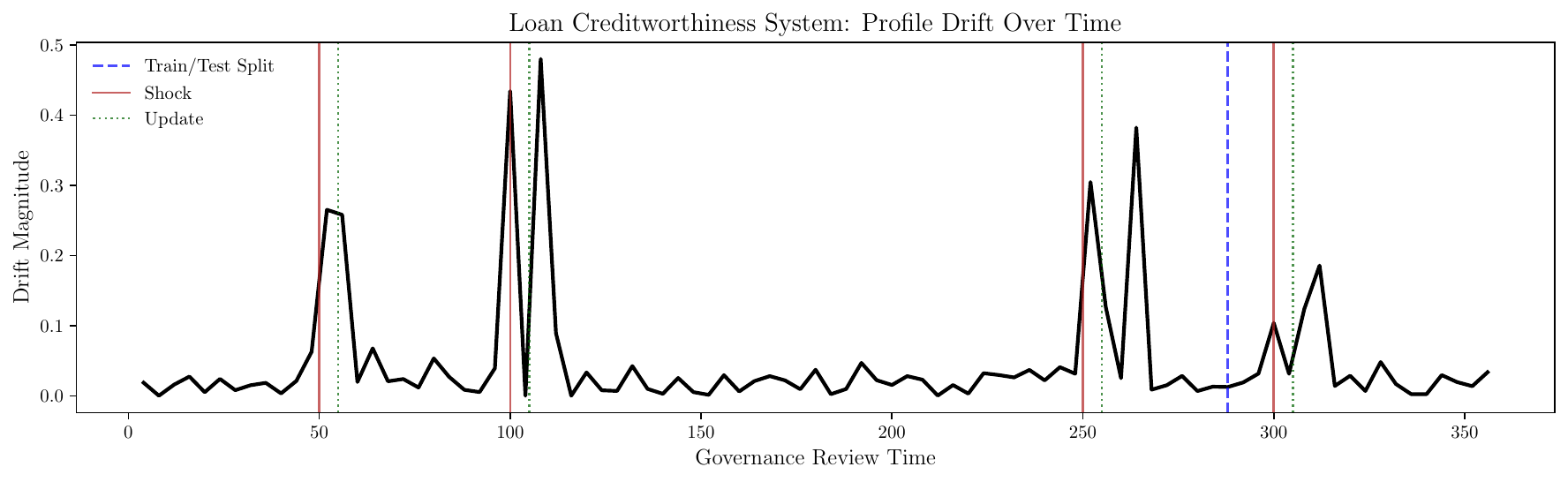}

\small (b) Profile drift over time

\caption{Scenario 2. Reporting boundary margin and profile drift over time.}
\label{fig:exp2_boundary_drift}
\end{figure}

This appendix provides the technical details of scenario~2, the single-system lifecycle baseline introduced in Section~\ref{subsec:exp2_main} and displays boundary margin and profile drift over time in Figure~\ref{fig:exp2_boundary_drift}.

\paragraph{Simulation design.}
Scenario~2 simulates a loan-creditworthiness AI system over \(360\) time steps. Four trustworthiness-relevant quantities are generated using the common simulation logic: performance, explainability, robustness, and fairness.
Performance, explainability, and robustness start from high baselines and gradually deteriorate. Fairness is simulated as an Equal Opportunity Gap \(g_{\mathrm{fair}}\), for which lower values indicate better fairness and which tends to widen over time in the absence of corrective updates. Before
construction of the trustworthiness profile, the fairness metric is re-oriented as
\[
Q_{\mathrm{fair}}(t)
=
1-g_{\mathrm{fair}}(t),
\]
so that all profile coordinates satisfy the common higher-is-better convention. Governance-review profiles are constructed at the union of all measurement times by carrying forward the latest available observation for
each dimension. Table~\ref{tab:exp2_design} summarizes the main design choices. For all four latent metric trajectories, we set \(\texttt{noise\_std}=0.01\). Thus, \(\varepsilon^{(1)}(t)\) has standard deviation \(0.005\), whereas
\(\varepsilon^{(2)}(t)\) has standard deviation \(0.01\). The transformation
from \(g_{\mathrm{fair}}\) to \(Q_{\mathrm{fair}}\) introduces no additional noise.

\begin{table*}[h!]
\scriptsize
\centering
\caption{Scenario~2: simulation design summary. Shocks occur at time steps \((50,100,250,300)\) and updates at \((55,105,255,305)\). Governance-review profiles are evaluated at the union of all metric timestamps, yielding \(90\) review points.}
\label{tab:exp2_design}
\begin{tabular}{l l r r l r r r}
\toprule
\textbf{Dimension} & \textbf{Raw metric} & \textbf{Baseline \(b\)} & \textbf{Drift \(d\)} & \textbf{Cadence} & \textbf{Obs.} & \textbf{Shock effects} & \textbf{Update effects} \\
\midrule
Performance    
& AUROC                       
& 0.95 & 0.20  
& every 4 steps  
& 90 
& \(-(0.25,0.40,0.05,0.05)\) 
& \((0.25,0.45,0.10,0.10)\) \\

Explainability 
& Reason-Code Stability       
& 0.95 & 0.18  
& every 12 steps 
& 30 
& \(-(0.10,0.20,0.10,0.05)\) 
& \((0.10,0.22,0.15,0.20)\) \\

Robustness     
& Stress-Test AUROC Retention 
& 0.95 & 0.10  
& every 12 steps 
& 30 
& \(-(0.15,0.25,0.25,0.05)\) 
& \((0.15,0.25,0.30,0.10)\) \\

Fairness       
& Equal Opportunity Gap       
& 0.05 & -0.30 
& every 16 steps 
& 23 
& \((0.15,0.15,0.05,0.05)\) 
& \(-(0.20,0.30,0.20,0.15)\) \\
\bottomrule
\end{tabular}
\end{table*}

\paragraph{Expert-based trustworthiness level rules.}
Table~\ref{tab:exp2_expert_rules} reports the expert-defined trustworthiness level rule used to generate the reference labels in scenario~2. Here, \(q_{\mathrm{perf}}, q_{\mathrm{exp}}, q_{\mathrm{rob}}\in[0,1]\) are normalized so that higher values indicate better quality, whereas \(q_{\mathrm{fair}}\in[0,1]\) is the lower-is-better Equal Opportunity Gap.

\begin{table}[h!]
\scriptsize
\centering
\caption{Scenario~2: reference trustworthiness level rule used to generate
the synthetic expert labels. All profile coordinates are oriented so that
higher values indicate greater trustworthiness.}
\label{tab:exp2_expert_rules}
\begin{tabular}{p{0.08\linewidth} p{0.62\linewidth} p{0.18\linewidth}}
\toprule
\textbf{Level} & \textbf{Rule} & \textbf{Interpretation} \\
\midrule

1 &
\(Q_{\mathrm{perf}} < 0.50\) or
\(Q_{\mathrm{fair}} < 0.65\)
& risky \\

2 &
otherwise
& acceptable \\

3 &
\(Q_{\mathrm{perf}} \ge 0.65\),
\(Q_{\mathrm{exp}} \ge 0.65\),
\(Q_{\mathrm{rob}} \ge 0.65\), and
\(Q_{\mathrm{fair}} \ge 0.75\),
provided that Level~4 does not apply
& good \\

4 &
\(Q_{\mathrm{perf}} \ge 0.75\),
\(Q_{\mathrm{exp}} \ge 0.75\),
\(Q_{\mathrm{rob}} \ge 0.75\), and
\(Q_{\mathrm{fair}} \ge 0.80\)
& excellent \\

\bottomrule
\end{tabular}
\end{table}

\paragraph{Learning results.} The selected candidate tree has three leaves. Up to the displayed numerical precision of the learned thresholds, it assigns level~3 when \(Q_{\mathrm{fair}}\leq 0.8067\), level~2 when \(Q_{\mathrm{fair}}>0.8067\) and \(Q_{\mathrm{exp}}\leq 0.6564\), and level~4 when \(Q_{\mathrm{fair}}>0.8067\) and \(Q_{\mathrm{exp}}>0.6564\). The resulting rule is not coordinatewise monotone. For example, consider two profiles that are identical except for fairness, with \(Q_{\mathrm{exp}}=0.60\), \(Q_{\mathrm{fair}}=0.80\) in the first profile, and \(Q_{\mathrm{fair}}=0.81\) in the second. The second profile weakly improves the first in every coordinate, yet the candidate assignment decreases from level~3 to level~2. Thus, there exist \(\mathbf q\preceq\mathbf q'\) such that \(\widehat{\mathsf{TW}}(\mathbf q)>\widehat{\mathsf{TW}}(\mathbf q')\). 
Because coordinatewise monotonicity is imposed as a structural requirement on trustworthiness level functions in Section~\ref{subsection:tw_level_functions}, this candidate fails the structural-admissibility condition at Gate~2 and is therefore not adopted; the expert-defined rule remains operative.

Table~\ref{tab:exp2_model_results} reports the predictive performance of the learned trustworthiness level rule in the asynchronous lending setting.
First, we note that the resulting reference-label distribution is highly imbalanced. Across the \(90\) governance-review profiles, \(78\) observations are assigned to level~4, while levels~1 and~2 occur three times each and level~3 occurs six times. Under the chronological \(80/20\) split, the training segment contains all four levels but only sparse support for levels~1,~2, and~3; the held-out segment contains only levels~3 and~4. On the training segment, the learned rule performs well, with low ordinal error (\(\mathrm{MAE}=0.11\)), near-perfect accuracy (\(0.96\)), and good balanced accuracy (\(0.75\)) and macro-F1 (\(0.71\)). Temporal cross-validation selected a shallow tree (\texttt{gini}, \texttt{max\_depth}=2, \texttt{max\_leaf\_nodes}=3, \(\texttt{ccp\_alpha}=0\)), which is consistent with the aim of learning a compact and auditable governance rule. 
Held-out performance is weaker than training performance: accuracy is \(0.83\), balanced accuracy is \(0.50\), macro-F1 is \(0.45\), and \(\mathrm{MAE}=0.17\). The discrepancy between ordinary and balanced accuracy reflects the strong imbalance of the reference levels and limited evidence for rarer regimes. The selected shallow tree therefore provides a compact candidate approximation of the dominant trustworthiness regime while
compressing distinctions among weakly represented levels. This illustrates why Gate~2 cannot rely on predictive accuracy alone: regime coverage, class
representation, interpretability, boundary stability, and preservation of
governance-relevant distinctions remain relevant to deciding whether the candidate should become operative.

\begin{table}[h!]
\centering
\caption{Scenario~2: predictive performance of the learned trustworthiness level rule. Temporal cross-validation selected a shallow tree configuration (\texttt{gini}, \texttt{max\_depth}=2, \texttt{max\_leaf\_nodes}=3, \(\texttt{ccp\_alpha}=0\)). Abbreviations: tr=train, te=test, Acc.=accuracy, Bal.Acc.=balanced accuracy.}
\label{tab:exp2_model_results}
\begin{tabular}{cccc}
\toprule
\textbf{MAE (tr/te)} & \textbf{Acc. (tr/te)} & \textbf{Bal. Acc. (tr/te)} & \textbf{Macro-F1 (tr/te)} \\
\midrule
0.11 / 0.17 & 0.96 / 0.83 & 0.75 / 0.50 & 0.71 / 0.45 \\
\bottomrule
\end{tabular}
\end{table}

%%%%%%%%%%%%%%%%%%%%%%%%%%%%%%%%%%%%%%%%%%%%%
% SINGLE-COLUMN FIGURE: BOUNDARY DISTANCE + PROFILE DRIFT

%%%%%%%%%%%%%%%%%%%%%%%%%%%%%%%%%%%%%%%%%%%
%%%%%%%%%%%%%%%%%%%%%%%%%%%%%%%%%%%%%%%%%%%
%%%%%%%%%%%%%%%%%%%%%%%%%%%%%%%%%%%%%%%%%%%
\subsection{Scenario 3}
\label{app:exp_3}
This appendix provides the technical details of Scenario~3, the AI deployment and governance-context comparison introduced in Section~\ref{subsec:exp3_main}.

\paragraph{Simulation design.}
Scenario~3 simulates two copies of the same clinical AI system over \(96\)
sparse monthly-like observations each. Hospital Alpha follows a slowly
degrading trajectory with no nonzero lifecycle-event effects, whereas
Hospital Beta experiences stronger degradation, adverse shocks, and
corrective updates. Performance, calibration, robustness, explainability, and
oversight quality are represented by higher-is-better profile coordinates.
Fairness is simulated as an Equal Opportunity Gap
\(g_{\mathrm{fair}}\), for which lower values indicate better fairness.
Before construction of the trustworthiness profile, it is re-oriented as
\[
Q_{\mathrm{fair}}(t)
=
1-g_{\mathrm{fair}}(t).
\]
All coordinates entering the trustworthiness profiles therefore satisfy the
common higher-is-better convention. Table~\ref{tab:exp3_design} summarizes the main design choices. For Hospital Alpha, we set \(\texttt{noise\_std}=0.004\), so the process-level perturbation has standard deviation \(0.002\) and the observation-level perturbation has standard deviation \(0.004\). For Hospital Beta, we set \(\texttt{noise\_std}=0.01\), so the corresponding standard deviations are \(0.005\) and \(0.01\), respectively.

\begin{table*}[h!]
\scriptsize
\centering
\caption{Scenario~3: simulation design summary. Both deployed copies are simulated over \(96\) sparse monthly-like observations. Hospital Alpha follows a slow-degradation trajectory with no nonzero event effects and is evaluated under a coarse local expert rule. Hospital Beta follows the common lifecycle simulation logic with signed lifecycle effects at time steps \((10,18,26,34,42,50,60,72)\). For higher-is-better dimensions, positive effects are corrective updates and negative effects are adverse shocks; for the fairness-gap coordinate, positive effects widen the gap and negative effects shrink it.}
\label{tab:exp3_design}
\begin{tabular}{l l r r r p{5.9cm}}
\toprule
\textbf{Dimension} & \textbf{Raw metric} & \textbf{Baseline \(b\)} & \textbf{Alpha drift \(d\)} & \textbf{Beta drift \(d\)} & \textbf{Hospital Beta signed lifecycle effects} \\
\midrule

Performance
& AUROC
& 0.93 & 0.035 & 0.12
& \((0.01,-0.20,0.10,-0.25,0.30,-0.15,0.25,-0.10)\) \\

Calibration
& Calibration Score
& 0.91 & 0.030 & 0.17
& \((0.01,-0.10,0.05,-0.10,0.25,-0.05,0.15,-0.25)\) \\

Robustness
& Stress-Test AUROC Retention
& 0.92 & 0.025 & 0.12
& \((0.01,-0.10,0.10,-0.15,0.25,-0.05,0.05,-0.15)\) \\

Fairness
& Equal Opportunity Gap
& 0.11 & -0.035 & -0.16
& \((-0.01,0.05,-0.05,0.05,-0.10,0.15,-0.25,0.15)\) \\

Explainability
& Reason-Code Stability
& 0.87 & 0.040 & 0.20
& \((0.01,-0.20,0.10,-0.15,0.30,-0.10,0.20,-0.25)\) \\

Oversight quality
& Oversight Quality
& 0.88 & 0.035 & 0.19
& \((0.02,-0.15,0.15,-0.20,0.25,-0.15,0.25,-0.15)\) \\
\bottomrule
\end{tabular}
\end{table*}

For Hospital Beta, the negative drift coefficient for the fairness coordinate means that the Equal Opportunity Gap tends to increase over time under the common simulation equation \(q(t)=\operatorname{clip}_{[0,1]}(b-d\tau_t+\cdots)\). Similarly, negative signed event effects on the fairness coordinate shrink the gap and therefore improve fairness, whereas positive signed event effects widen the gap and therefore worsen fairness.

\paragraph{Expert-based trustworthiness level rules.}
Hospital Alpha and Hospital Beta use the same ordered four-level governance scale, but differ in the granularity of their expert-defined trustworthiness level rules. Hospital Alpha uses a coarse local rule under which all observed profiles remain in level~4. This reflects a stable deployment environment in which slow degradation does not challenge the locally accepted highest trustworthiness regime. Hospital Beta uses a more refined rule, reported in Table~\ref{tab:exp3_expert_rules}, which adapts the threshold conditions to the six-dimensional clinical trustworthiness protocol. The Beta rule deliberately combines hard failure conditions on core dimensions with stricter conjunctive conditions for the upper trustworthiness regimes. This makes the learning problem more demanding than in the previous experiments while preserving the assumption that governance-relevant trustworthiness judgments can be represented through auditable threshold conditions.

\begin{table}[h!]
\scriptsize
\centering
\caption{Scenario~3: reference trustworthiness level rule for Hospital Beta.
All profile coordinates are oriented so that higher values indicate greater
trustworthiness.}
\label{tab:exp3_expert_rules}
\begin{tabular}{p{0.08\linewidth} p{0.66\linewidth} p{0.16\linewidth}}
\toprule
\textbf{Level} & \textbf{Rule} & \textbf{Interpretation} \\
\midrule

1 &
\(Q_{\mathrm{perf}} < 0.58\) or
\(Q_{\mathrm{cal}} < 0.56\) or
\(Q_{\mathrm{rob}} < 0.56\) or
\(Q_{\mathrm{fair}} < 0.50\)
& risky \\

4 &
\(Q_{\mathrm{perf}} \ge 0.85\),
\(Q_{\mathrm{cal}} \ge 0.85\),
\(Q_{\mathrm{rob}} \ge 0.85\),
\(Q_{\mathrm{fair}} \ge 0.85\),
\(Q_{\mathrm{exp}} \ge 0.60\), and
\(Q_{\mathrm{over}} \ge 0.75\)
& excellent \\

3 &
\(Q_{\mathrm{perf}} \ge 0.65\),
\(Q_{\mathrm{cal}} \ge 0.65\),
\(Q_{\mathrm{rob}} \ge 0.75\),
\(Q_{\mathrm{fair}} \ge 0.75\),
\(Q_{\mathrm{exp}} \ge 0.45\), and
\(Q_{\mathrm{over}} \ge 0.55\),
provided that Level~4 does not apply
& good \\

2 &
otherwise
& acceptable \\

\bottomrule
\end{tabular}
\end{table}

Hospital Alpha uses a coarser local rule: profiles are assigned level~4
unless a broad hard-stop condition is violated, namely
\(Q_{\mathrm{perf}}<0.70\),
\(Q_{\mathrm{cal}}<0.68\),
\(Q_{\mathrm{rob}}<0.70\),
\(Q_{\mathrm{fair}}<0.60\),
\(Q_{\mathrm{exp}}<0.55\), or
\(Q_{\mathrm{over}}<0.55\),
in which case they are assigned level~3. In the simulated Alpha trajectory,
none of these hard-stop conditions is reached. Hospital Beta occupies all four regimes, with \(31\) observations at level~4, \(37\) at level~3, \(20\) at level~2, and \(8\) at level~1.

\paragraph{Learning results.}
As in the previous scenarios, the data are split chronologically into a training segment containing the first (80\%) of observations and a test segment containing the remaining (20\%). For Hospital Alpha, no candidate trustworthiness level function is learned. All observed profiles remain in level~4 under the local expert-defined rule, so a supervised classifier would collapse to a constant assignment and add no governance-relevant information. Under Gate~1, empirical learning is therefore unwarranted and the expert-defined rule remains operative. For Hospital Beta, temporal cross-validation selects a shallow decision tree with \(3\) leaves (\texttt{gini}, \texttt{max\_depth}=2, \texttt{max\_leaf\_nodes}=3, \(\texttt{ccp\_alpha}=0\)). The learned rule for Hospital Beta depends only on performance: it assigns level~2 when \(Q_{\mathrm{perf}}\leq 0.6899\), level~3 when \(0.6899<Q_{\mathrm{perf}}\leq 0.8495\), and level~4 when \(Q_{\mathrm{perf}}>0.8495\). It is coordinatewise monotone. Increasing \(Q_{\mathrm{perf}}\) can only preserve or increase the assigned trustworthiness level, while changes in the remaining coordinates leave the assignment unchanged. Thus, unlike the candidate in Scenario~2, non-monotonicity is not the reason for rejection at Gate~2. However, monotonicity alone is insufficient for governance adequacy: the candidate does not recover the rare level~1 regime and reduces the
six-dimensional expert-defined rule to thresholds on performance alone, thereby discarding distinctions represented by calibration, robustness, fairness, explainability, and oversight quality. Table~\ref{tab:exp3_model_results} reports the predictive performance of all rules.

\begin{table*}[h!]
\small
\centering
\caption{Scenario~3: predictive performance of the operative or learned trustworthiness level rule. Hospital Alpha uses the expert-based level-\(4\) assignment because its local rule and trajectory contain no observed regime variation. Hospital Beta uses a learned decision tree selected by temporal cross-validation. Abbreviations: tr=train, te=test, Acc.=accuracy, Bal.Acc.=balanced accuracy.}
\label{tab:exp3_model_results}
\begin{tabular}{lcccc}
\toprule
\textbf{Scope} & \textbf{MAE} & \textbf{Acc.} & \textbf{Bal. Acc.} & \textbf{Macro-F1} \\
\midrule
Hospital Alpha expert rule & N/A& N/A& N/A& N/A\\
Hospital Beta (tr) & \(0.11\) & \(0.89\) & \(0.75\) & \(0.68\) \\
Hospital Beta (te) & \(0.50\) & \(0.50\) & \(0.50\) & \(0.33\) \\
\bottomrule
\end{tabular}
\end{table*}

The reference Hospital Beta labels are distributed across all four levels, whereas the learned rule predicts only levels~2,~3, and~4, with predicted counts \(18\), \(47\), and \(31\), respectively. This means that the rare level~1 regime is not recovered by the selected shallow tree. The deterioration on the held-out segment further shows that a compact learned rule may fail to preserve rare but important low-trustworthiness states when the lifecycle trajectory is unstable and the most severe regime is weakly represented.
Held-out performance also deteriorates substantially, with accuracy and balanced accuracy both equal to \(0.50\) and macro-F1 equal to \(0.33\). These results
provide evidence against adoption at Gate~2, so the expert-defined rule remains operative.

Candidate-rule diagnostics reinforce this conclusion. Hospital Alpha has small but nonzero profile drift, with mean drift \(0.0058\) and maximum drift \(0.0103\), reflecting very slow degradation within the locally accepted highest regime. Boundary margins under a learned tree are not reported for Hospital Alpha because no learned rule is used. Hospital Beta has mean candidate-rule boundary margin \(0.0595\), with
minimum margin \(0.0007\), showing that some lifecycle profiles lie very close to candidate regime transitions. Hospital Beta also exhibits substantially greater profile drift, with mean drift \(0.0261\) and maximum drift \(0.2402\). These values reflect the signed lifecycle events and resulting movement of the brittle deployment through trustworthiness space, especially in the later held-out segment where trustworthiness levels become less stable.

%%%%%%%%%%%%%%%%%%%%%%%%%%%%%%%%%%%%%%%%%%%%%
% FIGURE: EXPERIMENT 3
%%%%%%%%%%%%%%%%%%%%%%%%%%%%%%%%%%%%%%%%%%%%%
\newpage
\begin{figure*}[h!]
\centering

\begin{minipage}[t]{0.8\textwidth}
    \centering
    \includegraphics[width=\linewidth]   {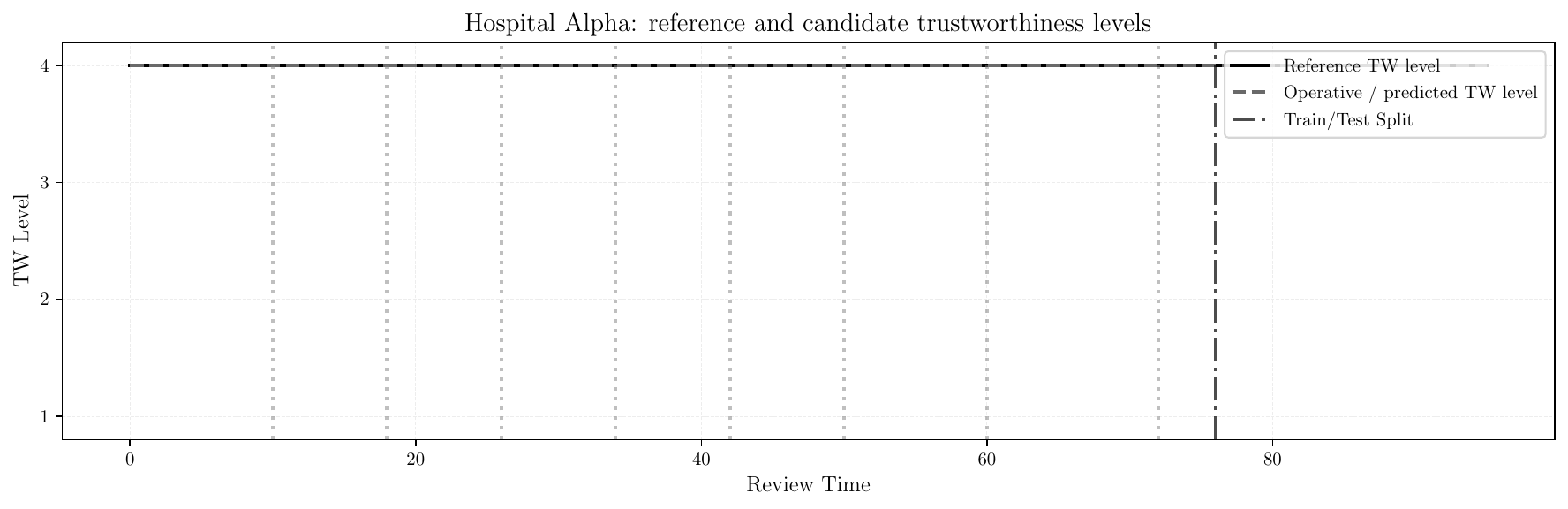}

    \smallskip
    \small
    (a) Hospital Alpha: expert-defined trustworthiness levels
\end{minipage}
\par\vspace{0.8em}
\begin{minipage}[t]{0.80\textwidth}
    \centering
    \includegraphics[width=\linewidth]
    {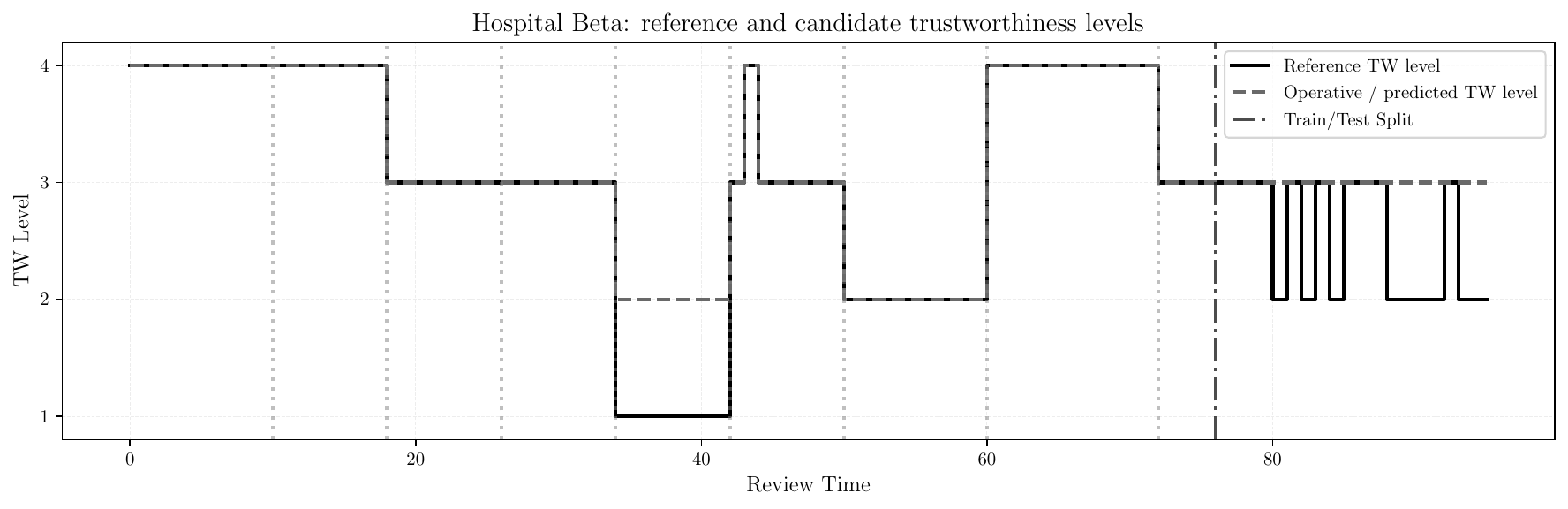}

    \smallskip
    \small
    (b) Hospital Beta: expert-defined and predicted
    trustworthiness levels
\end{minipage}

\caption{
Scenario 3: AI deployment and governance-context comparison.
The two panels compare simulated copies of the same clinical
decision-support system. Hospital Alpha remains in the highest
trustworthiness regime under a coarse local expert rule, whereas
Hospital Beta moves across multiple regimes under a more refined
expert-defined rule and lifecycle events involving adverse shocks
and corrective updates.
}
\label{fig:exp3_main_results}
\end{figure*}

%%%%%%%%%%%%%%%%%%%%%%%%%%%%%%%%%%%%%%%%%%%%%
% SINGLE-COLUMN FIGURE: BOUNDARY DISTANCE + PROFILE DRIFT
\newpage
\begin{figure}[h!]
\centering

%========================
% Panel (a): Boundary distance
%========================
\includegraphics[width=0.8\columnwidth]{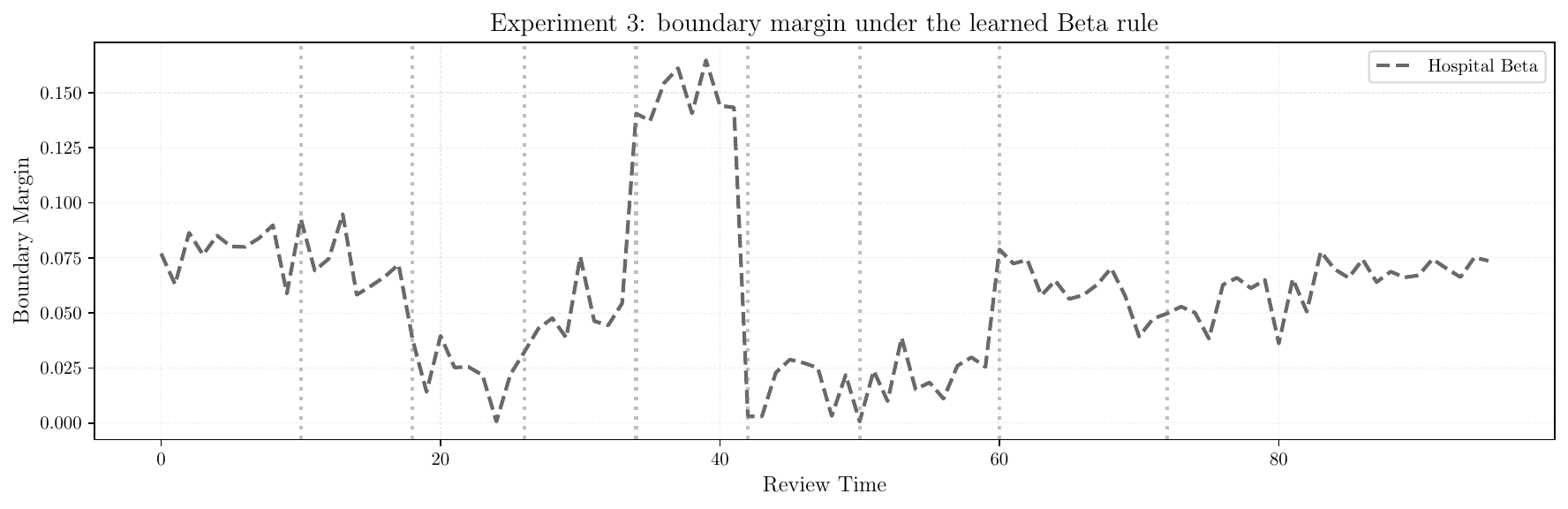}

\small (a)  Hospital Beta: Boundary margin over time

\vspace{0.6em}

%========================
% Panel (b): Profile drift
%========================
\includegraphics[width=0.8\columnwidth]{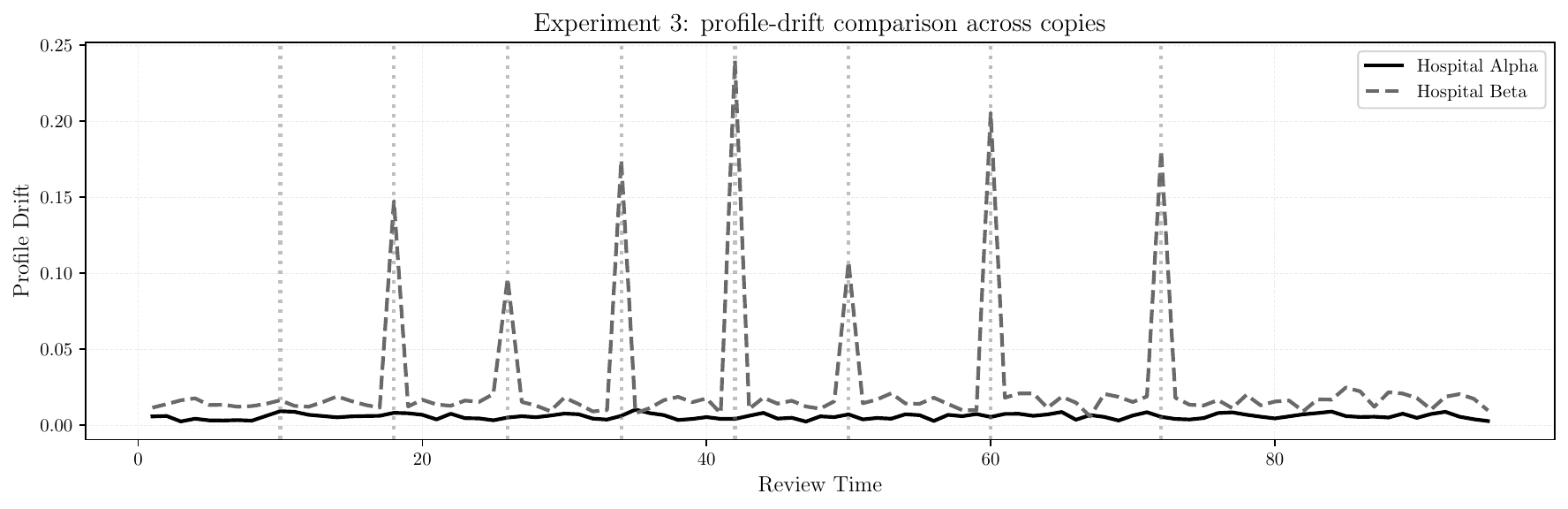}

\small (b) Profile-drift comparison across deployed copies

\caption{Scenario 3. Candidate-rule boundary margin for Hospital Beta and profile-drift comparison across the two deployed copies}
\label{fig:exp3_boundary_drift}
\end{figure}

%\bibliographystyle{ACM-Reference-Format}
\bibliographystyle{plainnat}
\bibliography{sample-base}

\end{document}